\documentclass{aa}

\usepackage{graphicx}
\usepackage{siunitx}
\usepackage{lipsum}
\usepackage{bm}
\usepackage{xcolor}
\usepackage{upgreek}
\usepackage{newtxtext}
\usepackage{anyfontsize}
\usepackage[varvw]{newtxmath}
\usepackage[shortlabels]{enumitem}
\usepackage{footmisc}

\defcitealias{whittingham2024}{W26}

\DeclareSymbolFont{bmisymbols}{OML}{cmm}{b}{it}
\DeclareMathSymbol{\bupsilon}{0}{bmisymbols}{"1D}

\newcommand{\rmn}[1]{\mathrm{#1}}
\newcommand{\bnabla}{\bm{\nabla}}
\newcommand{\bcdot}{\bm{\cdot}}
\newcommand{\bs}[1]{\boldsymbol{#1}}

\usepackage{makecell}

\graphicspath{{Figures/}}

\usepackage{xpatch}
\usepackage{hyperref}
\hypersetup{
	colorlinks=true,
	breaklinks=true,
	citecolor=blue,
	allcolors=blue,
	frenchlinks=true
}

\makeatletter
\xpatchcmd\NAT@citex
{%
	\@citea\NAT@hyper@{%
		\NAT@nmfmt{\NAT@nm}%
		\hyper@natlinkbreak{\NAT@aysep\NAT@spacechar}{\@citeb\@extra@b@citeb}%
		\NAT@date
	}%
}
{%
	\@citea
	\NAT@nmfmt{\NAT@nm}%
	\NAT@aysep\NAT@spacechar
	\NAT@hyper@{\NAT@date}%
}
{}{}
\xpatchcmd\NAT@citex
{%
	\@citea\NAT@hyper@{%
		\NAT@nmfmt{\NAT@nm}%
		\hyper@natlinkbreak{\NAT@spacechar\NAT@@open\if*#1*\else#1\NAT@spacechar\fi}%
		{\@citeb\@extra@b@citeb}%
		\NAT@date
	}%
}
{
	\@citea
	\NAT@nmfmt{\NAT@nm}%
	\NAT@spacechar\NAT@@open\if*#1*\else#1\NAT@spacechar\fi
	\NAT@hyper@{\NAT@date}%
}
{}{}
\makeatother

\makeatletter
\renewcommand*\aa@pageof{, page \thepage{} of \pageref*{LastPage}}
\makeatother

\begin{document}

\title{Zooming in on radio relics — II. How relic morphology probes density fluctuations at the edge of galaxy clusters}
\titlerunning{Zooming in on radio relics II.}

\author{Joseph Whittingham\inst{1}
    \and
    Christoph Pfrommer\inst{1}
    \and
    Maria Werhahn\inst{2}
    \and
    L\'{e}na Jlassi\inst{1,3}
    \and
    Philipp Girichidis\inst{4}
}

\institute{Leibniz-Institut f\"{u}r Astrophysik Potsdam (AIP), 
    An der Sternwarte 16, D-14482 Potsdam, Germany\\
    \email{jwhittingham@aip.de}
  \and
    Max-Planck-Institut f\"{u}r Astrophysik , Karl-Schwarzschild-Str. 1, 85748 Garching, Germany
  \and
    Institut für Physik und Astronomie, Universität Potsdam, Karl-Liebknecht-Str. 24/25, 14476 Potsdam, Germany
  \and
    Universit\"{a}t Heidelberg, Zentrum f\"{u}r Astronomie, Institut f\"{u}r Theoretische Astrophysik, Albert-Ueberle-Str. 2, 69120 Heidelberg, Germany \label{ITA}
}

\date{\today}

\abstract{Gas properties in the outer intracluster medium (ICM) are not well-constrained, as traditional probes lose sensitivity at Mpc distances. We show that the morphology of radio relics effectively encodes the power spectrum of the surrounding density fluctuations, and that they hence represent a new observational window. To demonstrate this, we use cosmologically motivated shock-tube simulations in which we systematically vary the coherence length, amplitude, and power-law slope of the upstream density power spectra. We then post-process our simulations with the cosmic ray electron spectral solver, \textsc{Crest}, thereby producing a suite of mock radio relics. We find that the downstream morphology of our simulated relics is independently sensitive to each of the aforementioned parameters. Specifically, we show that observed `double strand' features can be formed by curved shock fronts in projection, and that the scale of these features maps directly to the fluctuation coherence length. Increasing the fluctuation amplitude, meanwhile, progressively lengthens the downstream extent of the relic, thus explaining why relics are observed to be broader than the idealised expectation. It also broadens the Mach number distribution across the shock, which simultaneously increases the integrated radio flux density and produces patchier emission (especially notable at $\mathcal{M} \lesssim 2$). Finally, steepening the power-law slope makes `double strand' features more likely, and additionally increases both the number of radio filaments oriented parallel to the shock front and their spacing. At higher Mach numbers ($\mathcal{M} \gtrsim 3$), steepening the power-law slope can further lead to the production of curved radio filaments, which trace large eddies. We apply our analysis to the Toothbrush and Sausage relics, and find evidence for a typical fluctuation coherence length of $\sim$500~kpc, a non-uniform amplitude, and power spectra that are steeper than a Kolmogorov-like scaling ($\delta < -5/3$).} 


\maketitle
%

\section{Introduction}
\label{sec:introduction}

Gas properties in the intracluster medium (ICM) are primarily constrained through X-ray surface brightness and spectroscopy observations \citep{mohr1999, fabian2006, churazov2012, sanders2020} and the Sunyaev--Zel'dovich effect \citep[SZE;][]{sunyaev1980, birkinshaw1999, planck2013, bender2016}. However, these techniques become increasingly ineffective at large cluster-centric radii. For example, X-ray surface brightness scales as $S_\rmn{X} \propto \int n_\rmn{e}^2 T_\rmn{e}^{1/2}\mathrm{d}s$, where $n_\rmn{e}$ is the electron number density and $T_\rmn{e}$ is the electron temperature; it therefore declines rapidly in the low-density cluster outskirts. The SZE is more robust to this effect, as it probes the integrated electron pressure, with the Compton $y$-parameter scaling as $y \propto \int n_\rmn{e} T_\rmn{e}\mathrm{d}s$, where $s$ is the path length along the line of sight. This technique is not without problems either, though, as at the periphery of galaxy clusters $y\sim 10^{-6}$, and hence confusion from the cosmic infrared background (CIB) starts to become problematic \citep[see discussion in][]{walker2019}. As a result, to maximise the signal-to-noise ratio, observations using the above methods are often stacked \citep[see, e.g.,][]{planck2013, walker2013}, thereby blurring out the contribution of individual clusters.

Galaxy clusters are also host to a range of radio phenomena, which may provide a complementary, but as yet underutilised, window into ICM conditions. Here, we investigate the possibility of using radio relics to probe the ICM at distances $\gtrsim 1$~Mpc from the cluster centre. Radio relics are extended ($\sim$1~Mpc) sources of emission often found at the outskirts of merging clusters. They have typical surface brightnesses on the order of $1 \,\upmu$Jy arcsec$^{-2}$ at 1.4~GHz, which allows for substantially higher signal-to-noise ratios for individual objects compared to the aforementioned techniques. Surveys suggest that approximately 10\% of all clusters host at least one radio relic \citep{jones2023} and over 60 have been observed so far \citep{wittor2021}. This number is also expected to increase significantly once the Square Kilometre Array becomes operational \citep{braun2019, lee2024}.

Relics exhibit steep integrated spectra with spectral indices having $\alpha_\nu < -1$, where flux density scales with frequency as $S_\nu \propto \nu^{\alpha_\nu}$. Emission is often highly polarised, with average polarisation fractions typically exceeding 20\% \citep[see examples given in][]{wittor2019}. Together, these properties are usually interpreted as evidence that relics are the result of synchrotron emission produced by (re-)accelerated cosmic ray (CR) electrons. In particular, relics are believed to arise from diffusive shock acceleration \citep[DSA; ][]{krymskii1977, axford1977, bell1978, bell1978b, blandford1978} taking place at low Mach number ($\mathcal{M} \lesssim 3 - 5$) cluster merger shocks (see review by \citealt{brunetti2014} and summary of developments given in \citealt{whittingham2024} -- hereafter, \citetalias{whittingham2024}). 

The large-scale (${\gtrsim} 1$~Mpc) morphologies presented by radio relics can be highly varied. These differences are perhaps best illustrated by the relic in Abell 2256 \citep{vanweeren2012b,owen2014}, the Sausage relic \citep{kocevski2007, vanweeren2010}, and the Toothbrush relic \citep{vanweeren2012}, which have irregular filamentary, arc-like, and linear forms, respectively. Simulations have shown that shock fronts can produce similar structures \citep{ryu2008, hoefft2008, hoeft2011, skillman2011, nuza2017, lee2024, lee2024b}. Indeed, some specific mechanisms for explaining relic morphology based on this principle have already been proposed; for example, the origin of the surprisingly straight Toothbrush relic \citep{brueggen2012} and the origin of so-called `wrong-way round' relics, in which the spectral gradient points towards the cluster centre rather than away from it \citep{riseley2022, boess2023b}.

Radio relics also exhibit a series of smaller-scale  morphological features, the origin of which has been more difficult to unearth. For example, as resolution improves, it has become apparent that radio emission in galaxy clusters is often filamentary in nature \citep[see, e.g.,][]{botteon2020b, brienza2021, knowles2022, giacintucci2022}. Radio relics are no exception to this, with several reports of filamentary structure now existing in the literature \citep{owen2014, vanweeren2017, diGennaro2018, rajpurohit2020, deGasperin2022}. These filaments have a range of sizes and orientations; in some relics intermittent radio structures exist parallel to the expected shock front \citep{rajpurohit2022}. Further downstream, however, filaments are more typically oriented away from the shock instead and can even be curved -- see, e.g., the `brush' section of the Toothbrush relic \citep{rajpurohit2020}. Thicker filamentary structures can also appear along the length of a relic, manifesting as `double strands' composed of `strand' and `knot' elements \citep{rajpurohit2018, raja2024}. The origin of such structures is still unknown, but theories have been put forward based on magnetic flux-tubes \citep[see, e.g.,][]{owen2014, rudnick2022}, variable electron acceleration efficiency \citep{wittor2019, rajpurohit2022}, and the underlying shock morphology \citep{rajpurohit2020}.

Furthermore, relics always exhibit some degree of intensity fluctuations. These are observed both close to the shock front, which is generally understood to be caused by the local Mach number distribution (\citealt{wittor2021}, \citetalias{whittingham2024}), and in the downstream \citep[see, e.g.,][]{pal2025, rajpurohit2025}, which is more difficult to explain. Taken to its extreme, this patchiness produces splintered morphologies, where the relic appears to form disconnected islands \citep{vanweeren2013, chatterjee2024}. The origin of such fluctuations is equally unclear, although inhomogeneous magnetic fields, CR electron distributions, and shock properties could all play a role.

Finally, the downstream extent exhibited by observed radio relics is also puzzling, with some being remarkably extended and others much narrower. Indeed, variations in the downstream extent can be observed between relics in the same cluster \citep{deGasperin2022} and even along the same example -- see, e.g., the `handle' and `brush' of the Toothbrush relic \citep{rajpurohit2020}. Variance along the same structure, in particular, makes it difficult to explain this phenomenon through projection effects alone. Additionally, relics are found to be systematically wider than the theoretical expectation, as predicted by one-dimensional shock models. This is true even given highly favourable assumptions \citep{jones2023} and points to physics beyond the idealised DSA scenario.

In \citetalias{whittingham2024}, we analysed the evolution of shock waves in cosmological simulations of galaxy cluster mergers. We found that, over the distance range relevant for radio relics, merger shocks can collide with accretion shocks \citep[see also][]{zhang2020}. This leads to the formation of a dense, shock-compressed sheet; bound at the front by the shock surface and at the back by a contact discontinuity. As this sheet propagates outwards, it passes through inhomogeneities in the gas density field. Through a series of cosmologically-motivated shock-tube simulations, we showed that this scenario could explain: i) the radio vs.\ X-ray Mach number discrepancy, ii) the origin of $\upmu$G-strength magnetic fields, and iii) the spectral index evolution of relics in colour-colour diagrams \citep[problems 3, 4, and 5, as described in Sect.~1 of][]{whittingham2024}. The mock radio relics we generated were highly sensitive to the initial upstream density conditions. Indeed, they exhibited several of the features discussed above: namely, filamentary emission both at the shock front and in the downstream, intensity variations, and a more extended downstream. Here, we probe the origin of these features by extending our original simulations, modifying the power spectra of the upstream density field.

We organise the main body of the paper as follows: in Sect.~\ref{sec:methodology} we recap our methodology and explain how we explore the density parameter space. In Sect.~\ref{sec:hydro-analysis}, we analyse the impact of the fluctuations on hydrodynamics, including the generation of downstream substructure (Sect.~\ref{subsec:hydrodynamic-substructure}), the production of filamentary morphologies due to shock corrugation (Sect.~\ref{subsec:shock-corrugation}), and the phenomenon of shock fragmentation (Sect.~\ref{subsec:shock-fragmenetation}). In Sect.~\ref{subsec:radio-emission}, we extend our analysis to mock radio emission maps; specifically, we analyse the impact of the fluctuation amplitude and power-law slope on our mock radio relics (Sect.~\ref{subsec:suite-at-150MHz}), as well as the temporal evolution of the mock relics and the manifestation of `double strand' features (Sect.~\ref{subsec:radio-filaments}). In Sect.~\ref{sec:discussion}, we discuss our interpretation in the context of alternative mechanisms and, finally, in Sect.~\ref{sec:conclusions} we summarise our main conclusions.

We present additional material in the Appendix, including: the impact of the power-law slope on shock corrugation (Appendix~\ref{appendix:power-law-filaments}), an application of our analysis to the Sausage relic (Appendix~\ref{appendix:sausage}), the impact of our models on the Mach number distribution (Appendix~\ref{appendix:mach-no-dist}), the effect of coherence length on our mock radio relics at low frequency (Appendix~\ref{appendix:mock-emission-coherence-length}), and an analysis of the resulting spectral flux density (Appendix~\ref{appendix:flux-density}). In the later, we include a discussion of the factors necessary to bring the overall intensity of our mock relics into agreement with observations.

\section{Methodology}
\label{sec:methodology}

We research our proposed relic formation scenario (see Sect.~\ref{sec:introduction}) using shock tubes, which can achieve spatial resolution orders of magnitude higher than that possible in cosmological simulations. As shown in \citetalias{whittingham2024}, the point at which the merger and accretion shocks collide can be modelled as a Riemann problem, in which density is kept constant, but pressure varies between the up- and downstream. Here, we extend the original initial conditions (see Sect.~2 of \citetalias{whittingham2024}) to allow for a varying upstream density field.

\subsection[Arepo]{\textsc{Arepo}}

The shock tubes are modelled using the moving-mesh code \textsc{Arepo} \citep{springel2010, Pakmor2016I, weinberger2020}. This solves the equations of ideal magnetohydrodynamics (MHD) with a second-order finite-volume Godunov scheme \citep{pakmor2011, pakmor2013} and an HLLD Riemann solver \citep{miyoshi2005}. Mesh-generating points move approximately with the flow, whilst cells are allowed to re-fine and de-refine, so that their gas mass stays within a factor of two of $m_\rmn{target} \approx 1.5 \times 10^4 \,\rmn{M}_\odot$. This results in sub-kpc resolution in the downstream. Mass flux between cells prevents the code from being truly Lagrangian. To recover Lagrangian trajectories, we therefore employ the use of tracer particles (\citealt{genel2013}; see Sect.~\ref{subsec:crest}).

The $\mathbf{\nabla}\bcdot\bs{B}=0$ condition is enforced with a Powell 8-wave divergence cleaner \citep{powell1999}. \citet{pakmor2013} find that the Powell scheme is more robust than the competing scheme proposed by \citet{dedner2002}. Moreover, \citet{whittingham2021} find that this implementation produces divergence errors that are neither spatially- nor temporally-correlated, even in highly dynamic flows. This is important for the accurate modelling of CR electron spectra and subsequent radio synchrotron emission (see Sect.~\ref{subsec:crest} and~\ref{subsec:crayon}, respectively).

We run our \textsc{Arepo} simulations with the on-the-fly \citet{schaal2015} shock-finder and the CR physics module\footnote{CR protons are dynamically unimportant in our simulations, but are coupled to our CR electron implementation, as discussed in Sect.~\ref{subsec:crest}.} implemented by \citet{Pfrommer2017}, where CR protons are treated as a fluid with adiabatic index $\gamma_\rmn{a}=4/3$. The Mach number at a shock is calculated using the Rankine-Hugoniot jump conditions, with appropriate modifications for CR pressure and acceleration:
\begin{equation}
    \mathcal{M}^2 = \left(\frac{P_2}{P_1} - 1 \right) \frac{x_\rmn{s}}{\gamma_\mathrm{a, eff}(x_\rmn{s} - 1)},
    \label{eq:mach-no}
\end{equation}
where $P_1$ and $P_2$ are the up- and downstream sums of gas and CR pressures, respectively, $\gamma_\mathrm{a, eff}$ is the effective adiabatic index, and $x_\rmn{s}$ is the density jump at the shock. To increase numerical stability\footnote{See additional guards given in Sect.~2.4 of \citetalias{whittingham2024}.}, we identify shocks only when $\mathcal{M} \geq 1.3$. The dissipated energy rate at the shock is:
\begin{equation}
    \dot{E}_\rmn{diss} = \varepsilon_\rmn{diss}A_\rmn{shock}\frac{\mathcal{M}c_\mathrm{s,1}}{x_\rmn{s}},
    \label{eq:shock-dissipated-energy}
\end{equation}
where $\varepsilon_\rmn{diss}$ is the difference between the post-shock energy density and the adiabatically compressed pre-shock energy density, $A_\rmn{shock}$ is the cell's shock surface, and $c_{\rmn{s},1}$ is the upstream sound speed \citep[see][for further details]{Pfrommer2017}. We simulate CR proton acceleration through DSA using 10\% of the dissipated thermal energy.

\subsection[Crest]{\textsc{Crest}}
\label{subsec:crest}

We model CR electrons spectrally using the semi-analytic\footnote{\textsc{Crest} attaches the analytic result to the spectrum when one cooling process dominates and calculates the result numerically otherwise. This reduces the need to subcycle when electron cooling timescales become shorter than the MHD timestep.} solver, \textsc{Crest} \citep{winner2019}. This is a post-processing code that uses the tracer functionality in \textsc{Arepo} to store necessary variables on the MHD timestep. In total, we use $2.9\times10^6$ Lagrangian tracer particles, distributed initially on a grid with a spatial resolution of 2.5~kpc (see Sect.~\ref{subsec:shock-tube}). We solve the Fokker-Planck equation in the Lagrangian frame, with a spectrum being assigned to each tracer. In this way, the tracers represent a sampling of the underlying CR electron spectral density field. To assign values to the simulation volume as a whole, we then treat the tracers as mesh-generating points and apply a Voronoi tessellation.

Throughout this work, we express momentum in its dimensionless form: $p = \tilde{p} / (m_\rmn{e} c)$, where $\tilde{p}$ is physical momentum and $m_\rmn{e}$ is the electron rest mass. We further assume an isotropised pitch-angle distribution, which allows us to work in one-dimensional momentum space, where the one- and three-dimensional distribution functions are related by $f^\rmn{1D} = 4 \pi p^2 f^\rmn{3D}$. We calculate CR electron spectra over a momentum range of $p=10^{-2} - 10^{8}$ using twenty logarithmically-spaced bins per decade. We find that this is more than sufficient for spectral convergence.

It is generally believed that re-acceleration of pre-existing non-thermal populations is required in order to produce observed surface brightnesses \citep{pinzke2013, vazza2014, botteon2020}. Despite this, we initially assign a purely thermal spectrum to the tracers. This is reasonable as: i) fossil electron populations are currently not well-constrained\footnote{Discussed further in Sect.~\ref{sec:discussion}.}, and ii) previous work has shown that re-acceleration with a fixed acceleration efficiency only affects the normalisation of the spectrum, and not the spectral indices at radio-emitting frequencies \citep{pinzke2013, winner2019}; i.e. it should not affect the relic morphology, as is critical for this work.

In our spectral modelling, we account for DSA with quasi-parallel magnetic obliquity \citep[see formulae in][]{pais2018, winner2020}, adiabatic changes, and cooling via Coulomb, bremsstrahlung, inverse Compton, and synchrotron losses. To model DSA at shocks, we inject a power-law spectrum at shock-surface and post-shock cells (see definitions given in Sect.~2.4 of \citetalias{whittingham2024}). This is done by attaching a power-law slope with
\begin{equation}
\alpha = -\frac{x_\rmn{s} + 2}{x_\rmn{s}-1}
\label{eq:spectral-slope}
\end{equation}
to the one-dimensional distribution function. Following results by \citet{caprioli2020}, we limit the injection slope to a maximum of $\alpha=-2.2$. Similarly, following results from recent particle-in-cell simulations (see references given in the introduction of \citetalias{whittingham2024}) we do not accelerate CR electrons below $\mathcal{M}_\rmn{crit} = 2.3$. We choose the minimum and maximum momenta for acceleration dynamically, and model a super-exponential cut-off to mimic catastrophic cooling losses (see formula presented in Sect.~2.5 of \citetalias{whittingham2024}). We use 1\% of the CR proton energy to normalise the spectra between these momenta -- or equivalently 0.1\% of the liberated thermal energy at the shock (see Eq.~\ref{eq:shock-dissipated-energy}).

For cooling, we assume that the gas is fully ionized and of primordial composition. This equates to a hydrogen mass fraction of $X_\rmn{H} = 0.76$, a mean molecular weight of $\mu = 0.588$, and an ionisation fraction of $x_\rmn{e} = 1.157$. To aid comparison with observations we assume a redshift of $0.2$, which is the approximate redshift of the Toothbrush and Sausage radio relics \citep{vanweeren2010, vanweeren2012}. The energy density of the cosmic microwave background (CMB) is therefore set to $\epsilon_\rmn{CMB} = 8.65 \times 10^{-13}$~erg~cm$^{-3}$ or, equivalently, a magnetic field strength of $B_\rmn{CMB} \approx 4.7$ $\upmu$G.

\subsection[Crayon+]{\textsc{Crayon+}}
\label{subsec:crayon}

To generate mock emission, we use the \textsc{Crayon+} code \citep{werhahn2021}. This is capable of converting CR electron spectra into instantaneous emission for a variety of non-thermal radiation processes. Here, we focus solely on the radio synchrotron emission. Full details are presented in Sect.~2.4 and associated appendices of \citet{werhahn2021b}, however, the main result ultimately follows \citet{rybicki1986}. That is, the synchrotron emissivity $j_\nu$ of a tracer is given by:
\begin{equation}
j_\nu = \frac{\sqrt{3}e^3 B_\perp}{m_\rmn{e} c^2} \int\limits_{0}^{\infty} f(p) F(\nu / \nu_c) \mathrm{d}p\propto B_\perp^{1-\alpha_\nu}\nu^{\alpha_\nu},
\label{eq:synchrotron-emissivity}
\end{equation}
where $e$ is the elementary charge,  $B_\perp$ is the projection of the magnetic field onto the plane perpendicular to the line of sight, and $F(\nu/\nu_\rmn{c})$ is the dimensionless synchrotron kernel, with $\nu_\rmn{c}$ being the (momentum-dependent) critical frequency\footnote{See further definitions in \citealt{werhahn2021b} and \citetalias{whittingham2024}.}.

The specific radio synchrotron intensity, $I_\nu$, at frequency $\nu$ is then obtained by integrating $j_\nu$ along the line of sight, accounting for the aforementioned Voronoi tessellation. Radio spectral indices, in turn, are calculated as:
\begin{equation}
    \alpha^{\nu_2}_{\nu_1} = \frac{\log_{10}(I_{\nu_2} / I_{\nu_1})}{\log_{10}({\nu_2 / \nu_1})},
\end{equation}
where $\nu_2>\nu_1$ so that $\alpha^{\nu_2}_{\nu_1}<0$ for a spectrum that decreases with increasing frequency.

\subsection{Shock-tube setup and density fluctuations}
\label{subsec:shock-tube}

\begin{figure}
    \centering
    \includegraphics[width=1.0\columnwidth]{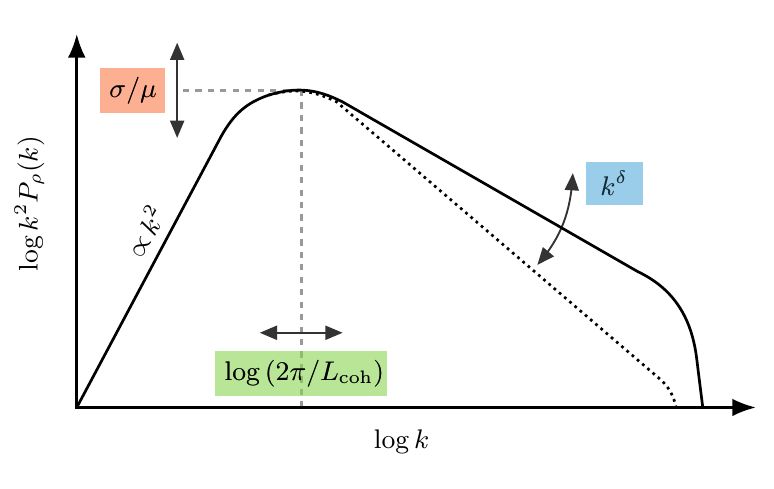}
    \caption{Schematic showing how we vary the 3D power spectra, $P_\rho$, of the upstream density fluctuations in our simulations. We probe the impact of the coherence length ($L_\rmn{coh}$), coefficient of variation ($\sigma/\mu$; i.e.\ amplitude), and the power-law slope ($\delta$). Power on spatial scales larger than the coherence length is set to white noise.}
    \label{figure:power-spectra}
\end{figure}

The shock tubes presented in this paper have periodic volumes of $1800\times300\times300$~kpc$^3$ and consist of four equal-sized sub-sections. The first and fourth regions serve purely to prevent interference from the reverse shock (see Sect.~2.3 of \citetalias{whittingham2024}). The second region, meanwhile, represents the high-resolution piston region, and the third contains our high-resolution initial upstream conditions. We define the start of this region to be $x=0$~kpc. 

Following analysis of cosmological simulations in \citetalias{whittingham2024}, we set the upstream pressure to $P_1 = 1 \times 10^{-13}$~dyne~cm$^{-2}$ and set the gas density across the whole box to be $\rho \approx 6.7 \times 10^{-29}$~g~cm$^{-3}$, or equivalently a thermal electron number density of $n_\mathrm{e} = 3.5 \times 10^{-5}$~cm$^{-3}$. This property is key to replicating the narrow, shock-compressed density sheet we observe in our cosmological simulations (see Sect.~3 of \citetalias{whittingham2024}). As shocks in observed radio relics cover a range of low Mach numbers \citep[see, e.g.,][]{vanweeren2017, wittor2021b}, we run all of our simulations in both $\mathcal{M}=2$ and $\mathcal{M}=3$ configurations. To do so, we set the initial downstream pressure to be $P_2 = 1 \times 10^{-12}$ dyne cm$^{-2}$ and $P_2 = 2.32 \times 10^{-12}$ dyne cm$^{-2}$, respectively.

\begin{figure}
    \centering
    \vspace{0.01cm}
    \includegraphics[width=1.0\columnwidth]{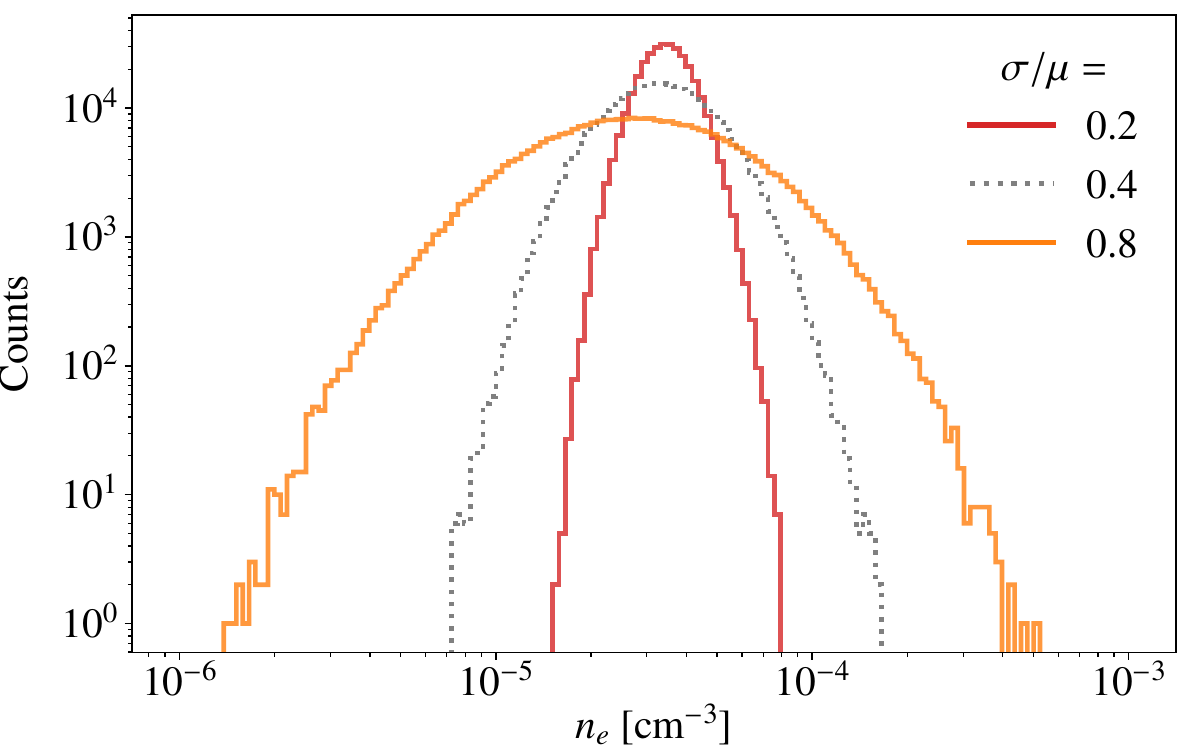}
    \caption{Histograms of the electron number density, calculated using all gas cells in our initial high-resolution upstream region. Each distribution is log-normal and has a mean density set to $3.5 \times 10^{-5}$~cm$^{-3}$. The standard value employed in our simulations is $\sigma/\mu=0.4$.}
    \label{figure:electron-density-pdf}
\end{figure}

There is now substantial evidence for the existence of density fluctuations\footnote{In \citetalias{whittingham2024}, we discussed these fluctuations in the framework of turbulence. Here, however, we remain more agnostic about their origin.} in the ICM; both in observations \citep{simionescu2011, eckhert2015, ghirardini2018} and in simulations \citep{nagai2011, zhuravleva2013, battaglia2015, angelinelli2021}. We add these to our high-resolution upstream region using the method of \citet{ruszkowski2007} and \citet{ehlert2018}. The advantage of this method is that the fluctuations are first created in $k$-space, allowing us precise control over their properties. 

Specifically, the power spectra can be altered in three main ways, as illustrated in Fig.~\ref{figure:power-spectra}:
\begin{itemize}
    \item Coherence length,  $L_\rmn{coh}$: this sets the typical wavelength of the perturbations. For $k < 2 \pi / L_\rmn{coh}$ we apply white noise; i.e.\ the power spectrum of the perturbations becomes scale-independent.
    \item Coefficient of variation, $\sigma / \mu$: this sets the amplitude of the density variations\footnote{For simplicity, we will generally refer to this as the `amplitude' in the remaining text.}.
    \item Power-law slope, $\delta$: this sets the relative amplitude of fluctuations on different scales, such that $k^2 P(k) \propto k^{\delta}$ between $k = 2 \pi / L_\rmn{coh}$ and the grid-scale. 
\end{itemize}

The true coherence length is relatively unconstrained. Here, we set the fiducial value to be $L_\rmn{coh} = 150$~kpc. This is half the smallest box dimension and the typical scale used in previous studies of relics \citep[see, e.g.,][]{dominguez-fernandez2021}. For the amplitude of fluctuations, we choose $\sigma / \mu = 0.4$, as was found to be typical for the periphery of merging clusters by \citet[][see in particular Figs.\ 2 and 4 therein]{zhuravleva2013}. We implement log-normal variance using a Box-Muller random variate method. Such distributions are inferred from both simulations and observations \citep[see, e.g.,][]{kawahara2008}. Finally, without further constraints, we choose a fiducial spectral slope of $\delta=-5/3$ (i.e.\ a \citealt{kolmogorov1941}-like slope). We explore the parameter space around these values, varying each parameter separately to investigate its impact on the mock relic morphology. In Fig.~\ref{figure:electron-density-pdf}, we present histograms of the different electron number density distributions implemented. We present a full list of the simulated density initial conditions in Table~\ref{tab:simulation_vars}. 

\renewcommand{\arraystretch}{1.2}
\begin{table}
\vspace{0.5cm}
    \centering
    \caption{The parameter space explored by our simulations, where each row indicates the upstream initial conditions in a given run.}
    \begin{tabular}{l || c | c | c}
\makecell{\textbf{Simulation} \\ \textbf{name}} & \makecell{\textbf{Coherence} \\ \textbf{scale [kpc]}} & \makecell{\textbf{Amplitude} \\ ($\sigma / \mu$)} & \makecell{\textbf{Power-law} \\ \textbf{slope}}  \\[8pt] \hline

$\rmn{Flat} $\,$\tablefootmark{(a)}$  & -      & 0            & -     \\  [5pt] 

$L_\rmn{coh}=75$~kpc  & 75      & 0.4            & -5/3     \\   
$L_\rmn{coh}=150$~kpc $\,$\tablefootmark{(b)} & 150     & 0.4            & -5/3     \\[5pt]

$\sigma/\mu = 0.2$    & 150     & 0.2            & -5/3     \\
$\sigma/\mu = 0.8$    & 150     & 0.8            & -5/3     \\[5pt]

$\delta=-2.5/3$       & 150     & 0.4            & -2.5/3   \\
$\delta=-10/3$        & 150     & 0.4            & -10/3    \\[5pt]
    \end{tabular}
    \label{tab:simulation_vars}
    \tablefoot{Each simulation is run both with a $\mathcal{M}=2$ and a $\mathcal{M}=3$ shock. \tablefoottext{a}{This simulation has no density fluctuations, as is often assumed in literature for theoretical calculations \citep[see, e.g.,][]{jones2023}.} \tablefoottext{b}{Our original `fiducial' setup, as previously analysed in \citetalias{whittingham2024}.}}
\end{table}

\subsection{Magnetic fluctuations}
\label{subsec:magnetic-fluctuations}

We include magnetic fluctuations in each simulation and implement them using the same process as above. Here, however, following \citet{ehlert2018}, we implement Gaussian variance, with the standard deviation of the distribution being fixed as a result of Parseval's theorem. We additionally fix the coherence length at 40~kpc. This is consistent with cosmological MHD simulations, which show that the magnetic power peaks in galaxy clusters on scales a factor of a few lower than the kinetic power \citep{tevlin2025}. We set the root-mean-square (RMS) strength to $\approx0.16$~$\upmu$G. This produces a mean plasma beta of $\langle P_\mathrm{th} / P_B\rangle = 100$ for the initial high-resolution upstream region, where $P_\mathrm{th}$ and $P_B$ are the thermal and magnetic pressures in each cell, respectively (see Appendix~E of \citetalias{whittingham2024}). This value is consistent with ICM simulations \citep{skillman2013, dominguez-fernandez2019, nelson2024, tevlin2025} and observations \citep{brunetti2001, govoni2017}. We remove any magnetic divergence from the initial conditions by projecting it out in $k$-space. To reduce complexity, we implement the same magnetic field initial conditions in every simulation; i.e.\ they are kept independent of the density fluctuations (see discussion in \citetalias{whittingham2024}). We will study the impact of magnetic fluctuations in future work.

\section{Hydrodynamic analysis}
\label{sec:hydro-analysis}

\subsection{Downstream substructure}
\label{subsec:hydrodynamic-substructure}

\begin{figure*}
    \centering
    \includegraphics[width=2.0\columnwidth]{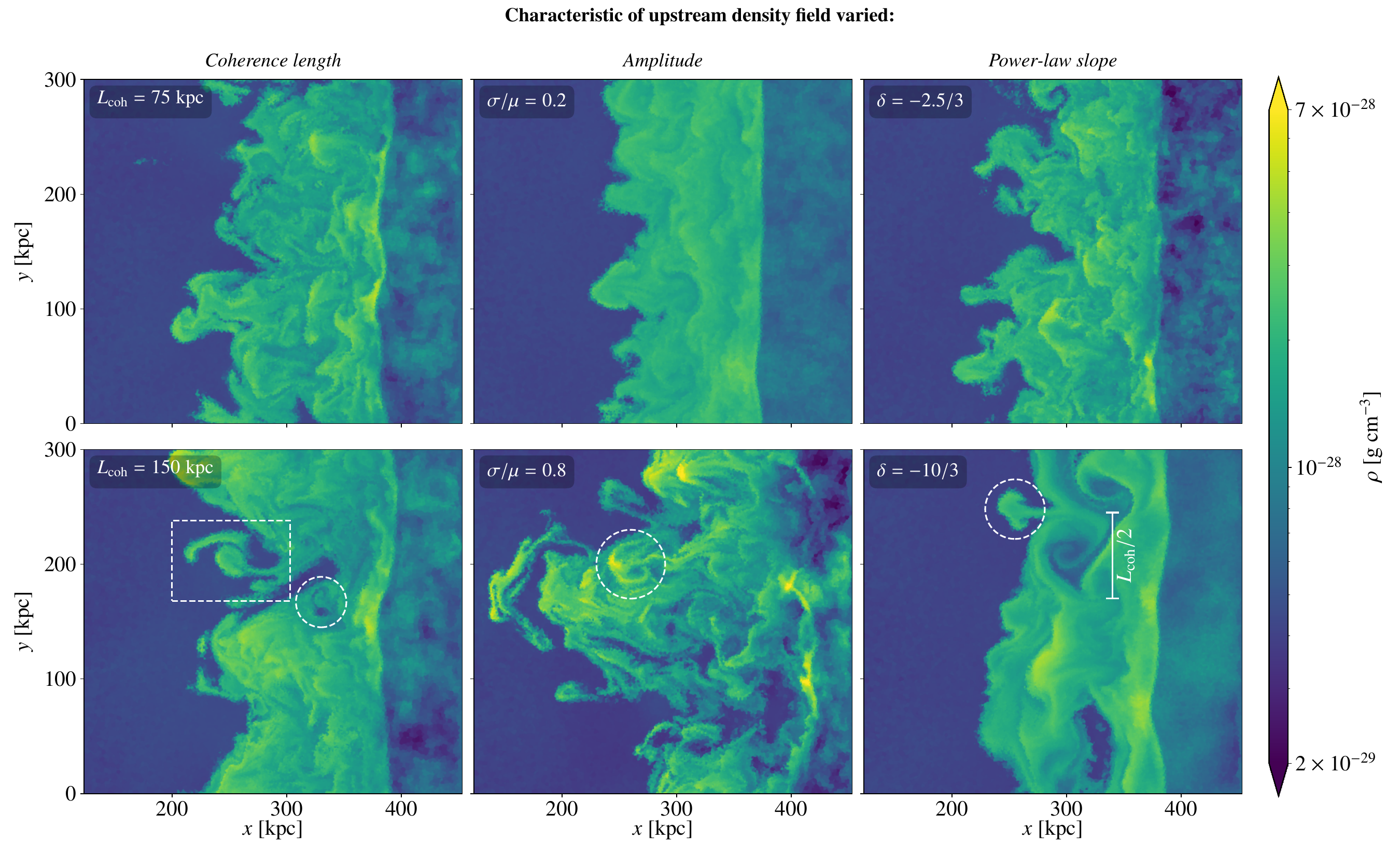}
    \caption{Slices through our $\mathcal{M}=3$ simulations at $t=244$~Myr, where the shock travels from left to right and colours indicate gas density. Dashed regions indicate secondary instabilities (see text). In each column, we vary only one characteristic of the upstream initial conditions (see Table~\ref{tab:simulation_vars}). Increasing the coherence length increases the width of the Rayleigh-Taylor `fingers'. Increasing the amplitude of the fluctuations extends the overall length of shock-compressed region and also enhances mixing. Finally, steepening the power spectra suppresses small-scale power, leading to smoother downstream density substructures but also the formation of large, undisturbed eddies.  An animated version of the figure can be found \href{https://youtu.be/oEtSXW41MwI}{here}.}
    \label{figure:density-slices}
    \vspace{-0.1cm}
\end{figure*}

We start our analysis by focussing on how different upstream models impact the development of gas substructure downstream. To this end, in Fig.~\ref{figure:density-slices} we show slices through the $\mathcal{M}=3$ simulations\footnote{The $\mathcal{M} = 2$ simulations have the same initial conditions but lower velocities, and hence generate less developed structures.} at $t=244$~Myr. At this point, the shock has almost left the high-resolution region. However, some indication of the initial upstream density conditions can still be seen on the right-hand side of each panel. From left to right, we vary the coherence length, amplitude, and the power-law slope of the density power spectra, respectively. It is immediately obvious that this has a significant impact on the downstream properties.

As discussed in \citetalias{whittingham2024}, substructure in the shock-compressed region forms primarily due to:
\begin{itemize}
    \item Inertia, as more massive density clumps are less easily accelerated by the shock.
    \item Vorticity, as induced by hydrodynamic instabilities and velocity shear.
    \item Compression and rarefaction, owing to varying downstream advection velocities.
\end{itemize}
The first and third points are a natural result of mass conservation ($\rho_1 \upsilon_1 = \rho_2 \upsilon_2$) combined with a fluctuating density field. The third point is also affected by vorticity, which allows the gas to be compressed along the $y$-~and $z$-axes as well.

In the left-hand column of Fig.~\ref{figure:density-slices}, we present simulations in which we vary only the coherence length of the upstream density field. This alters the characteristic length scale of the perturbations. We start with the $L_\mathrm{coh} = 150\,\mathrm{kpc}$ simulation, as the effects are easier to observe in this case. Our simulation box is 300~kpc tall. Hence, for $L_\mathrm{coh} = 150\,\mathrm{kpc}$, we expect to find two major overdensities and two underdensities along a given line of sight\footnote{\label{fn:animated-note}Seen particularly well in the animated version of the figure \href{https://youtu.be/oEtSXW41MwI}{here}.}. Indeed, the two overdensities lead to the formation of two `fingers' in the downstream. These are seeded by inertia, as described above, and are then further amplified by a Rayleigh–Taylor instability (see Sect.~4.1 of \citetalias{whittingham2024}). In addition, two secondary instabilities can be found in this simulation: firstly, a parasitic Kelvin-Helmholtz instability, which feeds off the generated velocity field and causes a vortex shedding pattern at $y \approx 200\,\mathrm{kpc}$ (see dashed rectangular region), and secondly, the formation of characteristic spiral eddies in the Rayleigh-Taylor troughs (dashed circular region)\footref{fn:animated-note}. Both of these increase the level of mixing downstream, further breaking the typical laminar-flow assumption applied to relics \citepalias[see Sect.~4.4 of][]{whittingham2024}. In the $L_\mathrm{coh} = 75\,\mathrm{kpc}$ run, the average separation between density peaks is substantially reduced. As a result, the induced vorticity fields produce overlapping motions, and the downstream morphology is consequently much less coherent.

In the middle column, we vary the amplitude of the initial fluctuations. This leads to the two main effects: i) a significantly extended downstream, and ii) an increase in small-scale substructure. The first effect is driven by a combination of factors. For example, with stronger underdensities, the shock is able to accelerate farther upstream\footnote{The shock front itself is also no longer as clearly delineated by a density jump compared to the other simulations. We will analyse this, however, in Sect.~\ref{subsec:shock-fragmenetation}.}. This stretches the shock-compressed region in the positive $x$-direction. Meanwhile, inertia means that the more massive overdensities are better able to resist acceleration, and hence better penetrate the downstream. This also results in a stronger corrugation of the contact discontinuity, thereby better seeding the Rayleigh-Taylor instability (see Sect.~B of \citetalias{whittingham2024}). This instability grows in the linear regime as $\exp(\omega t)$, with:
\begin{equation}
    \omega = \sqrt{|\bs{a}| \left( \frac{\rho_2 - \rho_3}{\rho_2 + \rho_3} \right) k},
    \label{eq:RTI-growth-rate}
\end{equation}
where $\bs{a} \propto -\bnabla P$ is the acceleration of the lighter material into the denser material, the density terms in the parentheses constitute the Atwood number, the subscripts `2' and `3' indicate post-shock gas and gas downstream of the contact discontinuity, respectively, and $k$ is the wave number of the perturbation \citep{chandrasekhar1961}. Increasing the amplitude of the fluctuations thus also increases the growth rate by producing a larger Atwood number, as well as increasing $|\bs{a}|$ in the Rayleigh-Taylor `troughs' (see Sect.~4.1 of \citet{whittingham2024}). This better reinforces the velocity field, thereby further extending the maximum distance reached by the shock-compressed material.

The small-scale substructure in the $\sigma/\mu=0.8$ simulation also promoted by a variety of factors. For example, varying inertia produces stronger regions of compression and rarefaction. Meanwhile, the increased development of the Rayeigh-Taylor instability results in significantly increased levels of mixing between gas in front of and behind the contact discontinuity. Indeed, the instability is active on all scales, and several smaller-scale Rayleigh-Taylor plumes can be seen aiding this process (see, e.g., the example highlighted in the dashed circular region in this panel, and the accompanying \href{https://youtu.be/oEtSXW41MwI}{movie}). On larger scales, the same effect brings low density gas, which was originally behind the shock-compressed region, right up to the shock front (see, e.g., at $y \approx 60\,\mathrm{kpc}$). Overall, this has the effect of increasing the inhomogeneity of the downstream density field. 

In the last column, we investigate the impact of the power-law slope. Making this steeper decreases the amplitude of fluctuations on small scales, which has the effect of smoothing the upstream density field (and the subsequent downstream compressed field). Conversely, making the slope shallower increases the amplitude of the small-scale fluctuations. The result is that, in the $\delta=-2.5/3$ simulation, the Rayleigh-Taylor instability is better seeded on small-scales, and goes non-linear faster. This reduces the coherence of the vorticity field, with similar results to the $L_\mathrm{coh} = 75\,\mathrm{kpc}$ run. In the $\delta=-10/3$ simulation, on the other hand, the suppression of small-scale power severely reduces the amplitude of large-$k$ fluctuations from which the Rayleigh-Taylor instability starts. The instability is subsequently unable to go non-linear within the simulation runtime, as it could in the previous simulations\footnote{A small Rayleigh-Taylor plume can be seen forming at $y \approx 240\,\mathrm{kpc}$ (as identified by the dashed circle in this panel), but this is now a secondary, parasitic instability.}. Instead, as is clear from inspecting the animated version of the figure, it is the Kelvin-Helmholtz instability that is dominant in this simulation. In this case, the shock front accelerates into the underdense regions, dragging downstream material with it. At the same time, overdense regions penetrate the downstream. The result is a strong, coherent velocity shear, which, in turn, gives rise to large, well-formed eddies. Indeed, the largest eddy scale is approximately 75~kpc, which is half the implemented coherence length (see barred white line in the panel). This is expected when there is a velocity shear on both sides of an underdensity.

\subsection{Large-scale filamentary structure}
\label{subsec:shock-corrugation}

\begin{figure*}
    \centering    \includegraphics[width=2\columnwidth]{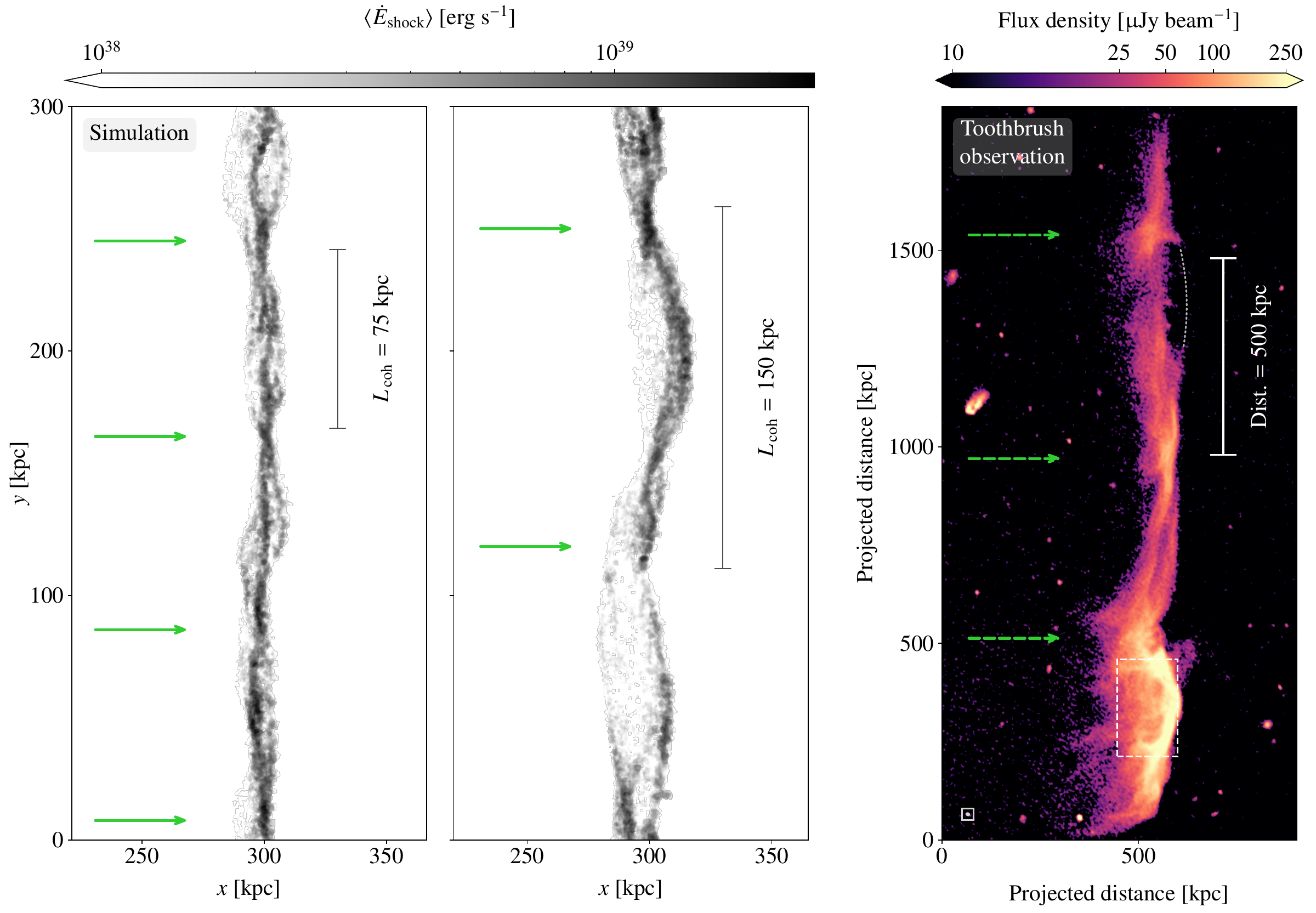}
    \caption{\textit{Left panel:} Shock-dissipated energy in our $\mathcal{M}=3$, $L_\mathrm{coh} = 75\,\mathrm{kpc}$ simulation at $t=190$~Myr, projected over a depth of 150~kpc. Green arrows indicate the rough position of `knots', where the morphology narrows. The coherence length of the upstream density fluctuations is indicated by the vertical barred line. \textit{Middle panel:} As previous, but for our $L_\mathrm{coh} = 150\,\mathrm{kpc}$ simulation. Both simulations otherwise have $\sigma/\mu=0.4$ and $\delta=-5/3$ (see Table~\ref{tab:simulation_vars}). \textit{Right panel:} VLA $L$-band 1–2~GHz high-resolution image of the Toothbrush relic. Data is from \citet{rajpurohit2020}. The beam size is shown in the bottom left corner. The dotted curved line represents a potentially missing part of the relic, discussed in Sect.~\ref{subsec:shock-fragmenetation}. The dashed rectangle, on the other hand, highlights an area with curved filaments, and is discussed in Sect.~\ref{subsec:suite-at-150MHz}. The projected shock surface produces `strand' and `knot' features that directly trace the upstream coherence length. If the elongated filamentary structures in the Toothbrush relic are also produced by this effect, the coherence length in the outer ICM would be $\sim$500~kpc. This is significantly larger than previously assumed.}
    \label{figure:shock-filaments}
\end{figure*}

As discussed in \citetalias{whittingham2024}, the acceleration of the shock into underdense regions results in the shock surface corrugating. On the left-hand side of Fig.~\ref{figure:shock-filaments}, we show the impact this has on the projected shock-dissipated energy for the $L_\mathrm{coh} = 75\,\mathrm{kpc}$ and $L_\mathrm{coh} = 150\,\mathrm{kpc}$ simulations. We have used a projected depth of 150~kpc in both cases, such that the left panel projects through $2 \times L_\mathrm{coh}$, whilst the middle panel projects only through $L_\mathrm{coh}$. We show how this can be used to constrain $L_\mathrm{coh}$ later in the section. The simulations are shown at $t=190$~Myr and are the $\mathcal{M}=3$ variations. The shock corrugation is particularly well-developed at this time, but the effects we discuss can be seen during the full simulation runtime (see discussion in Appendix~\ref{appendix:power-law-filaments}).

When seen edge-on, the curved shock surface produces an intricate, highly filamentary morphology. This includes several substructures that are qualitatively similar to those exhibited by observed radio relics. For example, using the terminology of \citet{rajpurohit2020}, we note the existence of `double strand' features, where a gap in dissipated energy forms between two arcing shock fronts. There are also bright `knots' or `twists', where the projected extent along the $x$-axis narrows. We have indicated these regions with green, horizontal arrows.

It has long been known that shock fronts can produce filamentary features in projection. Indeed, this has already been acknowledged in the radio relic community \citep[see, e.g.,][]{rajpurohit2020}. What has not been appreciated, however, is that, because the shock front corrugates on the scale of a typical upstream density fluctuation, these features may be a direct tracer of the coherence length of the upstream density field. In the left-hand panels, we indicate the size of the implemented coherence length as a barred line. It can be seen that these match the spacing of the arrows within a factor of a few percent.

On the right-hand side we show the VLA~$L$-band 1--2~GHz image of the Toothbrush relic, as initially presented in \citet{rajpurohit2020}. We have applied a power-law stretching of the colour map to emphasise the filamentary features. To help aid comparison, we have also rotated the image 65 degrees, compared to its on-the-sky position, and have converted the declination and right ascension to physical distances, assuming the relic is located at a redshift of $0.225$ \citep{vanweeren2012}. The true orientation of the Toothbrush along the line of sight is not known, and hence these are projected distances. It is, however, likely that we are observing the relic close to edge-on \citep{rajpurohit2022}.

We indicate three knots in the Toothbrush relic, each spaced roughly 500~kpc apart. Assuming that the filamentary radio morphology traces the underlying shock front\footnote{\label{fn:alternative-mechanism} See extension of our analysis to mock radio relics in Sect.~\ref{subsec:radio-filaments} and discussion of alternative mechanisms given in Sect.~\ref{sec:discussion}.}, this suggests a typical coherence length of the same size. This is significantly higher than the value commonly assumed by the community, and over three times the value used in previous idealised simulations of relics \citep{dominguez-fernandez2021, dominguez-fernandez2024}. In Appendix~\ref{appendix:sausage}, we repeat the same analysis for the Sausage relic \citep{kocevski2007, vanweeren2010}. Although the filamentary features are less distinct in this example, we find remarkably similar spacing here as well. 

Subsonic turbulence is often invoked to explain ICM fluctuations. However, it seems unlikely that it could maintain fluctuations with a coherence length of 500~kpc; developed turbulence requires a few eddy turnover times to form \citep{ryu2008}, and given an eddy-turnover speed of 350~km~s$^{-1}$, as is common in our simulations, the turnover time at the outer scale would be $ \tau = {\pi\, 500\,\rmn{kpc}}/{350\,\rmn{km}\,\rmn{s}^{-1}} \approx 4.4 \,\rmn{Gyr}$. Hence, establishing a classical $\delta = -5/3$ spectrum via driven turbulence across the inertial range would require a substantial fraction of the age of the Universe. Indeed, it would require longer than the age of the cluster even given higher turnover speeds.

If the large-scale filamentary structure in the Toothbrush relic is caused by shocks in projection, the separation of the `double strands' also provides additional information. For example, if the power-law slope of the density power spectrum is too shallow, small scale fluctuations will become more apparent and the `double strand' feature will lose coherence -- indeed, we show this explicitly in Sect.~\ref{subsec:radio-filaments} and Appendix~\ref{appendix:power-law-filaments}. The fact that the `double strands' in the Toothbrush are well-separated (as can be seen between $y=500$~kpc and $y=1,000$~kpc in Fig.~\ref{figure:shock-filaments}) consequently suggests that the upstream power-law slope in this case is relatively steep ($\delta \lesssim -5/3$).

A coherent `double strand' feature similarly requires the relic to have a depth of between $\sim L_\rmn{coh}$ and $\sim2 \times L_\rmn{coh}$. This is because as the projected depth increases, so does the number of fluctuations probed with a given line of sight. These are generally independently positioned, and hence the overall coherence of the feature will reduce again. On the other hand, if the depth is too far below the coherence length, there will be no overlapping shock fronts at all. To help show this effect, we provide an animated version of Fig.~\ref{figure:shock-filaments} \href{https://youtu.be/X1qEgLY4f8o}{here}, where we progressively increase the projected depth. Following this analysis, our earlier estimate for the coherence length implies that the Toothbrush relic has a structural depth of between $\sim500$~kpc and $\sim1$~Mpc along the line of sight. This is well within the virial size of its host cluster \citep{itahana2015}

Finally, for coherent `double strands' to form, the shock fronts in projection must be separated by a distance greater than the typical downstream extent -- if not, radio emission will fill in the gap. Indeed, this is likely happening in the case of Sausage relic, as analysed in Appendix~\ref{appendix:sausage}. Clearly, `strand' features will thus be more likely when the downstream emission is narrower. This can take place for a mixture of reasons, including: when cooling is more efficient, the downstream advection is slower, when the relic has not been forming for very long, and, simply, at higher frequency observations. Indeed, we examine the impact of these last two factors in Sec.~\ref{subsec:radio-filaments}. Alternatively, the gap between strands must be increased. Physically, this depends on the ability of the shock front to accelerate into underdense regions, which in turn depends on the typical size of an underdensity (itself dependent on the coherence length and the power-law slope) and the amplitude of the fluctuation. The former sets the maximum distance that the shock front can move ahead before encountering overdense material, whilst the later sets the speed at which it is able to accelerate into the region.

\subsection{Shock fragmentation}
\label{subsec:shock-fragmenetation}

\begin{figure*}
    \centering
    \includegraphics[width=2.\columnwidth]{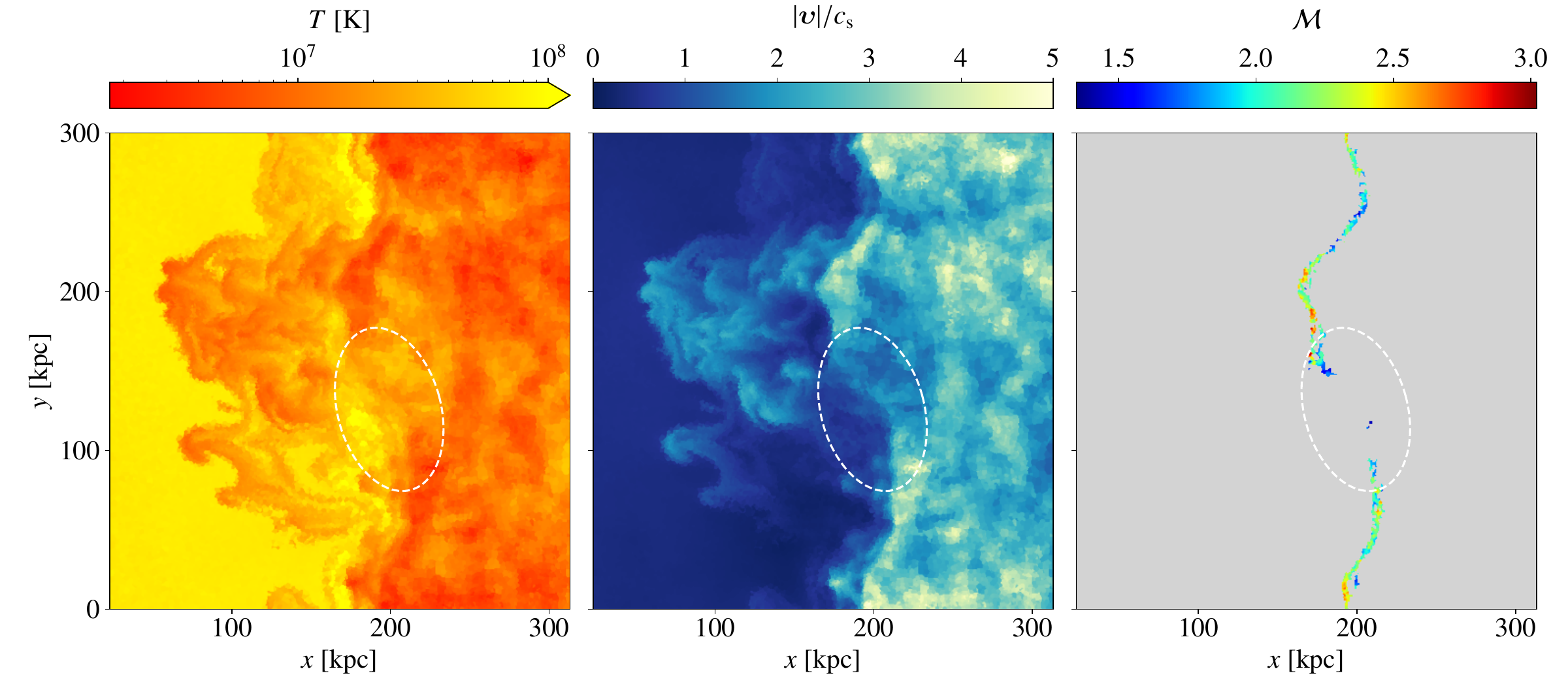}
    \caption{\textit{Left panel:} Slice showing gas temperature in our $\mathcal{M}=2$, high fluctuation amplitude ($\sigma/\mu = 0.8$) simulation at $t=170$~Myr. \textit{Middle panel:} As previous, but showing shock-frame gas speed divided by sound speed (see text). \textit{Right panel:} Thin projection of the emission-weighted Mach number with depth 35~kpc. Grey indicates no shock found. In our simulations, higher temperatures have higher sound speeds, which reduces the local shock strength. At its most extreme, this effect causes gaps to form in the shock surface, as it temporarily transforms into a pure sound wave (region indicated by the dashed white ellipse). This mechanism could help produce disjointed relics.}
    \label{figure:shock-gaps}
\end{figure*}

In the right-hand panel of Fig.~\ref{figure:shock-filaments}, we indicate a partially-formed `double strand' feature with a vertical, curved, dotted line; here, the filament starts on both sides, but is missing in the middle. Upstream fluctuations can also produce such a feature: in our simulations, the density fluctuations are in pressure equilibrium. This means that they are balanced inversely by temperature fluctuations; i.e.\ denser regions are colder and vice versa. These temperature fluctuations map directly to sound speed fluctuations \citepalias[see also Sect. 4.2.2 of][]{whittingham2024}. Given sufficiently underdense regions, the upstream sound speed can thus become comparable with the local shock speed\footnote{A corollary to this statement is that the Mach number distribution becomes broader given stronger density fluctuations. We show this to be true in Appendix~\ref{appendix:mach-no-dist}.}. At this point, the shock wave transforms into a pure sound wave, thereby fracturing the shock front.

We show an example of this process in Fig.~\ref{figure:shock-gaps}, marking the rough area where the shock fragments with a dashed white ellipse in all panels. We use our $\mathcal{M}=2$, high fluctuation amplitude ($\sigma/\mu=0.8$) simulation. This has the lowest average shock speed (${\sim}1,000\,\rmn{km}\,\rmn{s}^{-1}$) and the highest upstream sound speeds, as a result of the broader density range (see Fig.~\ref{figure:electron-density-pdf}). In the left-hand panel of Fig.~\ref{figure:shock-gaps}, we show the gas temperature. The underdense upstream regions are significantly hotter than their surroundings, and have comparably higher sound speeds. It can be seen that, in the white, dashed region, the temperature jump is minimal, as is expected at low, or even subsonic, Mach numbers. 

In the middle panel, we show the shock-frame gas speed divided by local sound speed. As in \citetalias{whittingham2024}, we calculate the shock-frame by subtracting the median shock speed from all cells. This is valid, as the shock only experiences minor velocity fluctuations along its surface relative to its average velocity. Naturally, this frame significantly overestimates the expected Mach number downstream, where gas is not moving at the shock speed. However, it provides a good approximation of how the local Mach number will evolve in the upstream. It can be seen that the hotter, underdense gas will produce lower Mach numbers, including below our numerical threshold of $\mathcal{M} = 1.3$. Finally, in the right-hand panel we show the true emission-weighted Mach number over a thin projection\footnote{As discussed in \citetalias{whittingham2024}, numerical broadening occasionally results in our shock finder not finding shock surface cells at a given hydro-timestep. These are, however, found immediately in the next timestep. By projecting through a thin region we mitigate this effect.} with 35~kpc. It can be seen that a sizeable chunk of the shock surface is missing, due to the aforementioned process.

The size of these gaps depends on the amount of sufficiently hot upstream gas. This is, naturally, tied to the coherence length and power-law slope, as these set the typical size of the fluctuations. For radio relics, the effect can also be exacerbated through the use of a critical Mach number, as more of the Mach number distribution will fail to meet the required threshold for acceleration (see Appendix~\ref{appendix:mach-no-dist}). Nonetheless, we note that unless a significant percentage of the line of sight falls below this threshold, some emission will still be visible.

We have previously shown that, as underdensities result in local acceleration, lower Mach numbers are statistically found in a more advanced position \citepalias[see Appendix~C of][]{whittingham2024}. As a result, it is more likely that a line of sight towards the front of the relic will probe Mach numbers below the threshold. This is consistent with data from the Toothbrush, where the more advanced part of the `strand' is missing. That is not to say, however, that the rear section of a `double strand' feature cannot also break; indeed, the middle panel of Fig.~\ref{figure:shock-filaments} shows that, given the right projection depth, this too is possible.

The Toothbrush is far from the only relic to exhibit fragmentation in this manner. Indeed, we show how the same effect may be at play in the Sausage relic in Appendix~\ref{appendix:sausage}. Moreover, there are now several examples of relics in the literature that are strongly fragmented, such that their overall morphology starts to appear patchy and irregular \citep[see, e.g.,][]{urdampilleta2018, zhang2020b, domingue-fernandez2026}. In these cases -- and especially in the cited examples -- the distance across the narrowest side tends to be far greater than that possible through downstream advection alone. This suggests that the relics are being viewed at an oblique angle. In our framework, this correlation is expected because oblique angles shorten the line-of-sight distance through the relic, making it more likely that only regions below the Mach number threshold are intersected.

\section{Mock radio emission}
\label{subsec:radio-emission}

We now extend our analysis to mock radio emission maps. As discussed in Sect.~\ref{sec:methodology}, we produce these using the comprehensive CR electron spectral modelling code \textsc{Crest}, with spectra post-processed in turn using the emission code \textsc{Crayon+}. In doing so, we make the minimum possible assumptions for the resultant radio emission.

\subsection{Impact of amplitude and power-law slope}
\label{subsec:suite-at-150MHz}

\begin{figure*}
    \centering
    \small\textbf{Mach~2}\par\medskip
    \includegraphics[width=2.0\columnwidth]{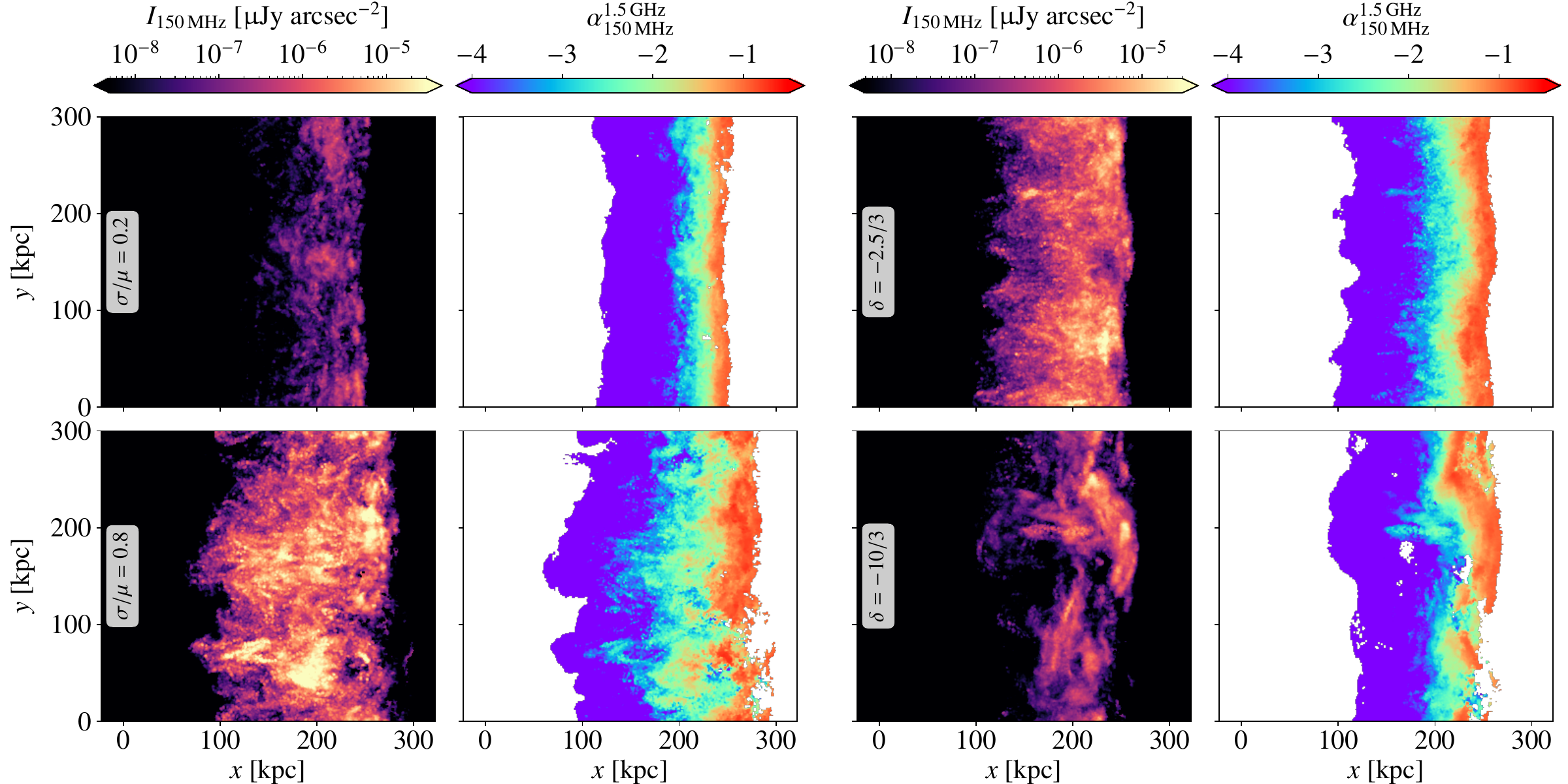}
    \caption{\textit{Left-most panels:} Synchrotron intensity maps at 150~MHz and spectral index maps taken between 1.5~GHz and 150~MHz for the Mach~2 high fluctuation amplitude ($\sigma / \mu$) variations at $t=244$ Myr. The projection depth is 300~kpc. We have masked spectral index values where no emission takes place along the line of sight. Increasing $\sigma / \mu$ increases the extent of the downstream emission and the average intensity. A varying amplitude of upstream fluctuations may explain the variable downstream extent exhibited by the Toothbrush relic (cf. right panel of Fig.~\ref{figure:shock-filaments}). \textit{Right-most panels:} As previously, but for the power-law slope variations. A steeper power-law slope produces more elongated filaments and causes a greater fraction to be oriented parallel to the shock front. We vary the coherence length in Appendix~\ref{appendix:mock-emission-coherence-length}. An animated version of the figure can be found \href{https://youtu.be/68OLoUN1IZE}{here}.
    }
    \label{figure:mach-2-variations}
\end{figure*}

\begin{figure*}
    \centering
    \small\textbf{Mach~3}\par\medskip
    \includegraphics[width=2.\columnwidth]{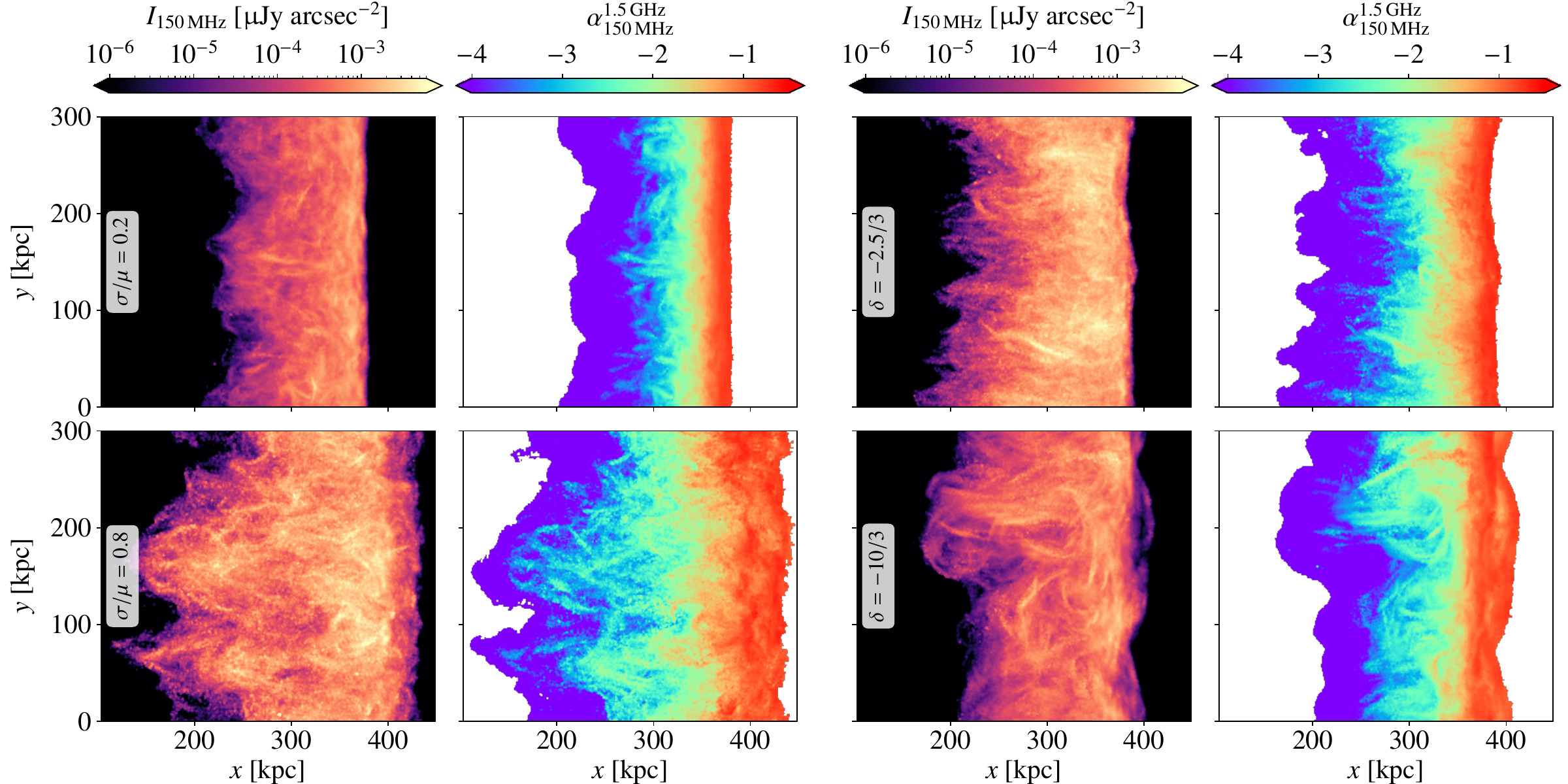}
    \caption{As Fig.~\ref{figure:mach-2-variations} but for our Mach~3 simulations. Emission is more extended in each case relative to the corresponding Mach~2 simulation. Increased mixing and less variation in electron spectral slopes produces lower average levels of intensity fluctuations. Filamentary emission is more prevalent, however, resulting from the better amplified magnetic field. In the $\delta=-10/3$ simulation, curved filaments are especially noticeable. These are produced by the eddies analysed in Fig.~\ref{figure:density-slices}. We note the existence of such curved filaments in the `brush' section of the Toothbrush relic (see dashed rectangle in Fig.~\ref{figure:shock-filaments}). An animated version of the figure can be found \href{https://youtu.be/T036RSr0va8}{here}.}
    \label{figure:mach-3-variations}
\end{figure*}

We start with analysis our Mach~2 simulations: in Fig.~\ref{figure:mach-2-variations}, we show the mock synchrotron intensity and spectral index maps for the amplitude ($\sigma/\mu$) and power-law slope ($\delta$) variations, respectively. Intensity maps are shown at 150~MHz, and the spectral index maps are calculated between 1.5~GHz and 150~MHz. Each simulation is shown at $t=244$ Myr and has a projection depth of 300~kpc. We analyse the coherence length ($L_\rmn{coh}$) variations separately in Appendix~\ref{appendix:mock-emission-coherence-length}, as we find only limited variation between our mocks at these low frequencies in the sampled parameter space.

In the left-hand panels of Fig.~\ref{figure:mach-2-variations}, we vary the amplitude of the upstream density fluctuations. This substantially increases i) the average intensity, ii) the level of variation in intensity -- especially close to the shock front, and iii) the downstream extent. These effects are all consistent with our previous hydrodynamical analyses. For example, in Appendix~\ref{appendix:mach-no-dist}, we show that increasing the amplitude of the fluctuations produces a broader Mach number distribution. This produces flatter electron spectra, which significantly boosts the radio emission\footnote{The radio emission is also slightly increased by a better amplified magnetic field, with root-mean-square values increasing by roughly a factor of two. This results in an increase in intensity by a factor of $\sim$4 (see Eq.~\ref{eq:synchrotron-emissivity}).} \citepalias{whittingham2024}. The result is that, compared to the $\sigma/\mu =0.2$ simulation, the $\sigma/\mu =0.8$ simulation is brighter\footnote{The overall intensity of the mock relics presented here is still much reduced compared to observed relics. We show in Appendix~\ref{appendix:flux-density}, however, that we can successfully match observed intensities once some additional factors are taken into account. These factors generally only affect the normalisation of the CR electron spectra, not its spectral slope, and thus do not affect the analysis performed in this section.} on average by a factor of roughly~100. Indeed, we show in Appendix~\ref{appendix:flux-density} that the boost compared to a simulation with a uniform upstream density ($\sigma/\mu =0$) is roughly~1000. 

The increased width of the Mach number distribution automatically produces patchier surface brightness maps (see Sect.~\ref{subsec:shock-fragmenetation}). This is accentuated by the increased inhomogeneity of the downstream density field (see Fig.~\ref{figure:density-slices}). As discussed in Sect.~\ref{subsec:hydrodynamic-substructure}, increasing the amplitude of the fluctuations results in a more variable downstream velocity field, as well as a better seeding of the Rayleigh-Taylor instability. The Rayleigh-Taylor instability, in particular, is thus better able to bring both aged electrons from the back of the shock-compressed region, as well as non-injected gas from behind the contact discontinuity, right up to the shock front. A good example of this effect is at $y\approx60$~kpc, as previously analysed in Fig.~\ref{figure:density-slices}. This causes substantial fluctuations in the surface brightness and spectral index maps in the $\sigma/\mu =0.8$ run in the same region.

It can be seen that the emission in the $\sigma/\mu = 0.8$ simulation extends roughly twice as far in the $x$-direction compared to the $\sigma/\mu = 0.2$ analogue. As previously analysed in Sect.~\ref{subsec:hydrodynamic-substructure}, this takes place due to a mixture of: i) higher inertia in more massive overdensities, ii) a better seeded Rayleigh-Taylor instability reinforcing this velocity field, and iii) the acceleration of the shock front into underdense regions stretching the relic in the positive $x$-direction as well. Observationally, it has been noted that relics are always more extended than the one-dimensional theoretical expectation, which assumes a steady-state relic with uniform up- and downstream gas conditions. This idealised scenario neglects density fluctuations, which provide a natural explanation for the varying downstream extents. 

Indeed, when the amplitude of the upstream density fluctuations varies in space, the downstream extent of the radio relic will vary as well. This could naturally explain the formation of the `handle' and `brush' in the Toothbrush relic (compare the $\sigma/\mu$ variations with the observation reproduced in Fig.~\ref{figure:shock-filaments}). This scenario would also be consistent with polarisation data, which shows that the `brush' is substantially less polarised compared to other parts of the relic \citep{kierdorf2017}. This is usually interpreted as being a result of Faraday depolarisation, implying that the `brush' section of the relic is located behind the cluster \citep[see, e.g.,][]{rajpurohit2020b, rajpurohit2022, hoeft2022}. In our scenario, however, at least some of the lower polarisation is intrinsic, resulting instead from increased velocity fluctuations driven by hydrodynamic instabilities, as analysed in Sect.~\ref{subsec:hydrodynamic-substructure}.

In the right-hand panels of Fig.~\ref{figure:mach-2-variations}, we vary the power-law slope of the density fluctuations. In the top row, the slope is shallower, and hence there is greater power on small-scales. In contrast, in the bottom row, the slope is steeper, and the density fluctuations are correspondingly smoother. In Fig.~\ref{figure:density-slices}, we showed that steepening the power-law slope produced large eddies downstream, with outer scales equal to half the coherence length; i.e.\ equal to $L_\rmn{coh} / 2 = 75$~kpc. Like the Rayleigh-Taylor instability, these eddies also act to mix old electron populations with freshly injected regions. This produces a morphology that is notably patchier than in the $\delta=-2.5/3$ case. When the eddies form too close to the shock front, they can disrupt it, as can be seen in both the intensity and spectral index maps at $y \lesssim 100$~kpc. The large eddies additionally cause shearing, which amplifies the magnetic field \citepalias{whittingham2024}. This produces elongated radio filaments, which are oriented roughly parallel to the shock front. In contrast, in the $\delta=-2.5/3$ simulation, filaments are oriented predominantly downstream, away from the shock.

In Fig.~\ref{figure:mach-3-variations}, we show the analogous plot for our Mach~3 variations. We have re-scaled the colourbar here by the average increase in emission, but have kept the same dynamic range. The shock speed is now approximately 1,500~km~s$^{-1}$ (compared to the previous value of 1,000~km~s$^{-1}$) and the post-shock speed is comparably higher. As a result, CR electrons are advected further downstream. This increases the average physical extent of the mock relics and ensures that fresher CR electrons with higher intensities are also found farther downstream, thereby keeping the relic brighter for a longer distance. 

The higher speeds additionally result in a higher level of mixing. This makes it less likely that an individual line of sight will pass only through low energy-density CR electrons, thereby reducing spatial intensity variation. At higher Mach numbers, spectral slopes also exhibit less variation, which additionally contributes to this effect \citepalias[see Sect.~4.3 of][]{whittingham2024}. Indeed, when comparing between all models, it is noticeable that the variation in average intensity is much reduced compared to the $\mathcal{M}=2$ simulations (we quantify this  statement in Appendix~\ref{appendix:flux-density}).

The increased downstream turbulence additionally leads to a more amplified magnetic field \citepalias{whittingham2024}. This produces stronger filamentary structures in each mock relic. In general, such structures are oriented away from the shock surface. By comparing between the runs where we have varied the power-law slope, it can be seen, once again, that steepening this slope is particularly influential for reorienting the filaments to be parallel to the shock surface. Moreover, it increases the spacing between the filaments, as is more typical for observed relic morphologies \citep[see, e.g.,][]{rajpurohit2022}.

In the downstream region away from the shock front, the $\delta=-10/3$ simulation additionally produces large, curved filaments. These are created by the shearing effect of the Kelvin-Helmholtz vortices previously analysed (see Sect.~\ref{subsec:shock-corrugation}). Such filaments are observed in real radio relics. For example, curved filaments are clearly present in the relic in Abell 2256 \citep{rajpurohit2022}. Moreover, we identify similar curved filaments in the `brush' section of the Toothbrush relic in Fig.~\ref{figure:shock-filaments} (see the region highlighted with a dashed white box). It is notable that out of all the simulations we present, only the $\delta=-10/3$ runs are able to generate this type of filament. This is thus further evidence for the existence of a smooth ($\delta \lesssim -5/3$) density field in the outer ICM of galaxy clusters.

Finally, we note the existence of apparent upstream emission in our simulations, as previously reported by \citet{lusetti2025}. This is especially obvious in the high fluctuation amplitude ($\sigma/\mu=0.8$) run, where emission peaks around $x\approx400$~kpc but extends towards $x\approx425$~kpc. It can also be seen, however, in the steep power-law slope ($\delta=-10/3$) simulation, although here the upstream emission is generally more filamentary. We will analyse the origin of this emission in future work.

\subsection{Initial evolution and `double strand' features}
\label{subsec:radio-filaments}

\begin{figure}
    \centering
    \includegraphics[width=1.\columnwidth]{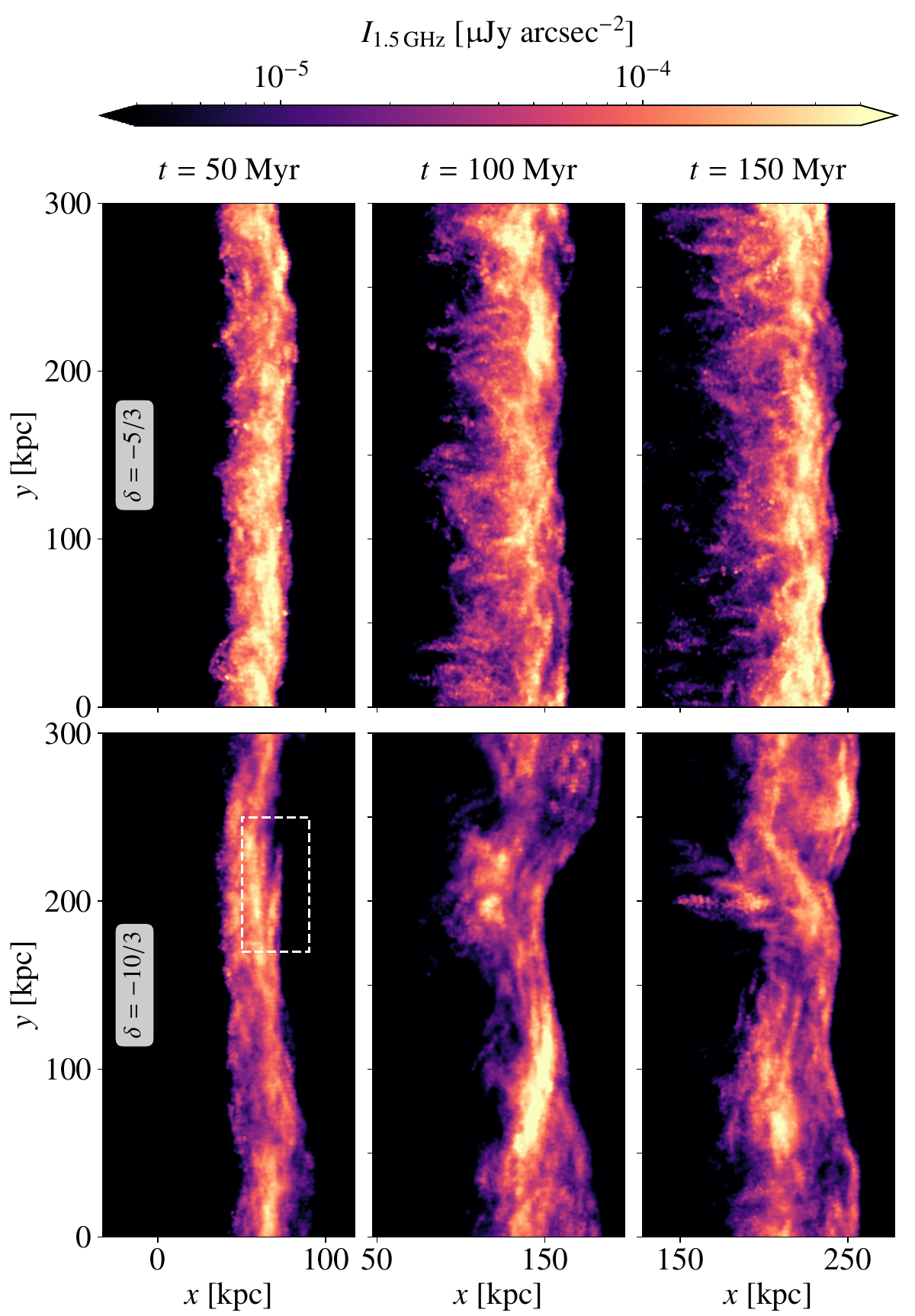}
    \caption{\textit{Top row:} Synchrotron intensity maps at 1.5~GHz for our Mach~3 $L_\rmn{coh}=150\,\rmn{kpc}$ simulation (with $\delta=-5/3$) shown at 50~Myr intervals. \textit{Bottom row:} as previous, but for our $\delta=-10/3$ simulation. The dashed white box indicates a radio spur (see text). In contrast to the standard paradigm, in our scenario the radio relic only starts forming when the merger shock collides with an accretion shock. This leads to an (initially) increasing downstream extent over time. Even at early times the properties of our mock relics compare favourably with observations. We note, however, that only the steeper power-law slope ($\delta=-10/3$) can produce coherent `double strand' features.}
    \label{figure:mock-relic-time-series}
\end{figure}

In the standard paradigm, radio relics form as a merger shock traverses a cluster. That is, the acceleration takes place from the moment the merger shock is created, with emission then building up downstream until radiative cooling establishes a steady-state. In the scenario we propose, however, merger shocks must first collide with an accretion shock in order to produce a relic\footnote{As accretion shocks only take place in the outer ICM, a corollary of this statement is that radio relics can only start forming in the outer ICM as well. This may help explain why radio relics are not observed close to the cluster centre \citep{vazza2012}.}. Our scenario thus implies the existence of a population of nascent radio relics, in which the downstream has not yet reached its full extent. To illustrate this, we show a time series in Fig.~\ref{figure:mock-relic-time-series}, with synchrotron intensity maps shown at 1.5~GHz for our Mach~3 $L_\rmn{coh}=150\,\rmn{kpc}$ and $\delta=-10/3$ simulations over 50~Myr intervals. For this figure, we label the $L_\rmn{coh}=150\,\rmn{kpc}$ simulation as $\delta=-5/3$ to emphasise the difference in the power-law slope between the two models (see Table~\ref{tab:simulation_vars}).

It can be seen that, even at $t=50$~Myr, the simulations produce structures that are qualitatively similar to observed radio relics. The $\delta=-10/3$ simulation, especially, compares favourably: for example, it exhibits a radio `spur' at the shock front at $y\approx200$~kpc, which also appears in the Sausage relic (compare the white, dashed boxes in Fig.~\ref{figure:mock-relic-time-series} and Fig.~\ref{figure:sausage}). The width of the mock relics is also similar to both the Sausage and Toothbrush relics, once observational resolution is taken into account: in particular, the intensity profile shows qualitative agreement with the large-scale filamentary structures in the Toothbrush relic (see Fig.~\ref{figure:shock-filaments}), which are only $~20$~kpc across at 1.4~GHz and exhibit a relatively rapid drop in surface brightness at their trailing edge.

Over time, the downstream extent in both simulations increases. However, the causes of this are different: in the $L_\rmn{coh}=150\,\rmn{kpc}$ ($\delta=-5/3$) simulation, hydrodynamic instabilities at the contact discontinuity are the leading factor. This produces small-scale filamentary structures oriented along the $x$-axis. Shock corrugation, in this case, has a more secondary impact, producing the bright large-scale filamentary structures oriented along the $y$-direction. In the $\delta=-10/3$ simulation, the situation is reversed, with shock corrugation providing the majority of the relic's breadth, and instabilities at the contact discontinuity acting only in a secondary capacity.

It is notable that out of all of our mocks, only the $\delta=-10/3$ simulation can form coherent `double strand' features at 1.5~GHz. This is primarily due to the impact of small-scale density fluctuations in simulations with $\delta\geq-5/3$  (see also Appendix~\ref{appendix:power-law-filaments}). As previously discussed in Sect.~\ref{subsec:shock-corrugation}, the coherence of strands can be increased by either reducing the extent of the downstream emission or increasing the distance along the $x$-direction over which the shock front corrugates. We have already noted that the true coherence length is likely closer to 500~kpc than the implemented value of 150~kpc. Adjusting this parameter would help open the gap between strands, and thus allow the $L_\rmn{coh}=150\,\rmn{kpc}$ simulation to form `double strand' features in projection. However, it would not change the impact of small-scale density fluctuations, which reduce how smooth these features are. It is clear that out of the two simulations, the filamentary structure identified in the Toothbrush relic better reflects the $\delta=-10/3$ simulation. Fig.~\ref{figure:mock-relic-time-series} hence provides further indirect evidence for the existence of a smooth density field in the outer ICM of galaxy clusters, characterised by a steeply decreasing density power spectrum.

\begin{figure}
    \centering
    \includegraphics[width=1.\columnwidth]{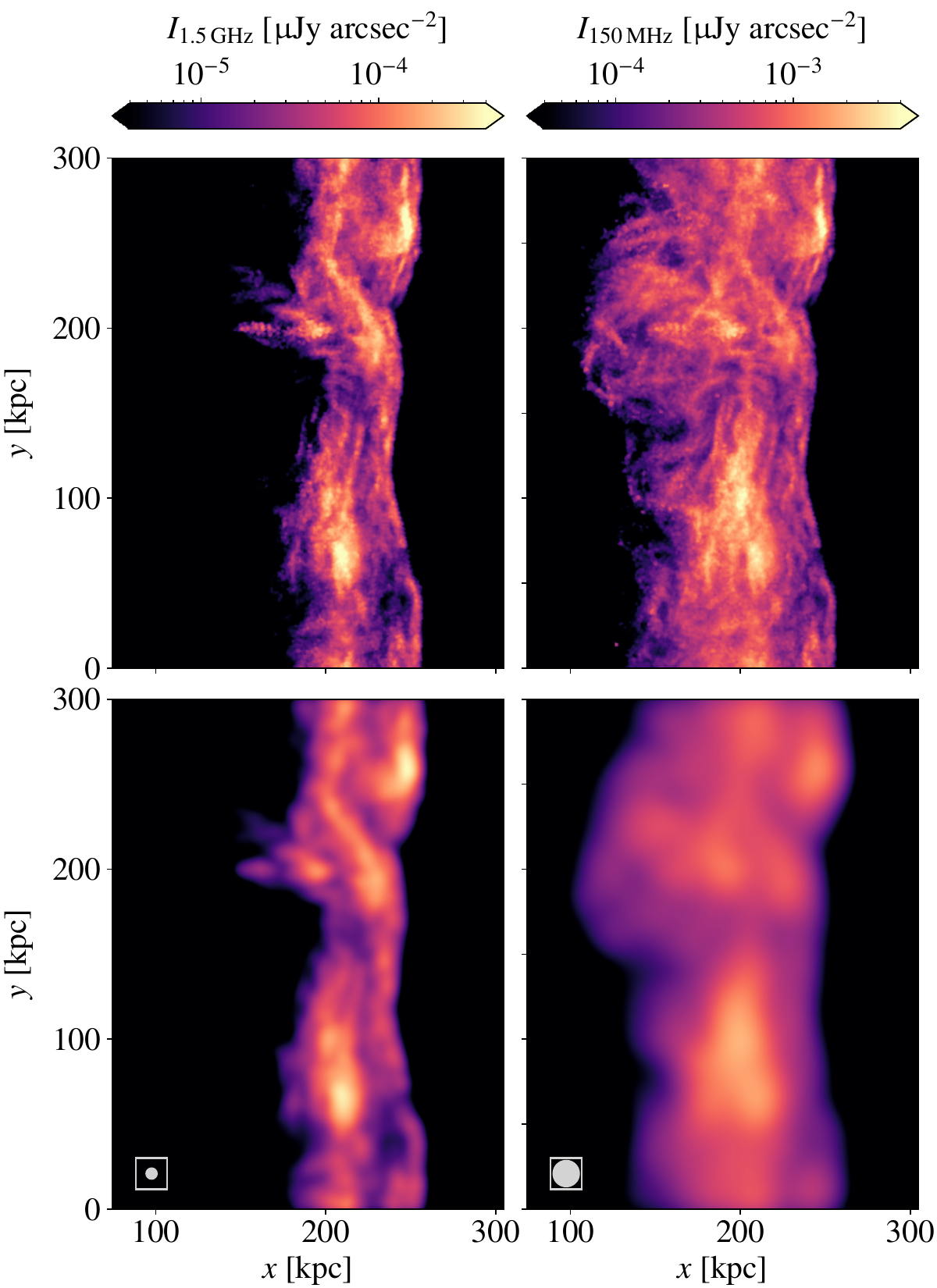}
    \caption{\textit{Top left:} Synchrotron intensity map at 1.5~GHz for our Mach~3 $\delta=$~$-10/3$ simulation at $t=150$~Myr. \textit{Top right:} As previous, but showing intensity at 150~MHz. \textit{Bottom panels:} Corresponding maps blurred with a Gaussian kernel to mimic current observational capabilities (see text). At lower frequencies, the downstream appears more extended, which reduces the coherence of `double strand' features. We predict that these features should also be visible at low frequencies, however, given better instrument resolution.}
    \label{figure:radio-filaments-vs-frequency}
\end{figure}

As discussed in Sect.~\ref{subsec:shock-corrugation}, large-scale filamentary structures should be more visible at high frequencies. This is because higher frequency intensity maps tend to probe higher CR electron momenta; these populations cool more quickly, leading to a more rapid drop-off in intensity and a correspondingly less extended downstream. In Fig.~\ref{figure:radio-filaments-vs-frequency}, we show this effect for the $\delta=-10/3$ simulation at $t=150$~Myr. We present the mock radio intensity map at 1.5~GHz in the top-left panel and at 150~MHz in the top-right panel. As expected, the coherence of the `double strand' feature is reduced at lower frequency, as radio emission better fills in the gap between individual strands. It is notable, however, that the strands themselves can still be picked out even at 150~MHz.

This is perhaps unexpected, given the general lack of observations of filamentary structures in relics at 150~MHz. The mock maps in the top row of Fig.~\ref{figure:radio-filaments-vs-frequency}, however, represent observations with perfect resolution. For a more apt comparison, in the bottom panels of Fig.~\ref{figure:radio-filaments-vs-frequency}, we have blurred the mocks using Gaussian kernels. Specifically, we have assumed the relics are located at a redshift of 0.2 (see Sect.~\ref{sec:methodology}) and have used a beam size with a full-width half-maximum (FWHM) of $1.7\arcsec$ and $4.3\arcsec$ for the 1.5~GHz and 150~MHz maps, respectively. These, in turn, represent the surface area\footnote{We have isotropised our kernel as our mocks have no preferential direction.} of the VLA $L$-band beam of 2.0\arcsec$\times$1.5\arcsec and of the LOFAR HBA beam of $4.8\arcsec\times3.9\arcsec$ \citep{vanweeren2016}, as are commonly used at these frequencies. 

It can be seen that, after a Gaussian blur has been applied, the large-scale filamentary structure in the 1.5~GHz mock is still very evident. Indeed, the structure compares favourably with that exhibited in the Toothbrush in Fig.~\ref{figure:shock-filaments}. On the other hand, the filamentary structure in the 150~MHz mock is much less clear, being effectively removed below $y\approx200$~kpc. Traces of the filamentary structure still exist above this point, and indeed particularly well-calibrated observations at 150~MHz are also able to pick out such structures (see, e.g., Fig.~4 of \citealt{vanweeren2016}). Figure~\ref{figure:radio-filaments-vs-frequency} shows, though, that LOFAR HBA observations at 150~MHz will generally struggle to observe such filaments. Whilst SKA1-Low's greater sensitivity will allow for the detection of fainter filaments, its angular resolution will not improve at this frequency. However, for sufficiently bright relics and well-defined `strands', imaging at 150~MHz at 1-2\arcsec resolution might be made feasible by incorporating international baselines with appropriate weighting \citep{ye2024}.

\section{Discussion}
\label{sec:discussion}

Density fluctuations are able to explain many morphological features present in real radio relics. However, they are not the only possible mechanism. Here, we discuss some alternatives and compare them.

Filamentary synchrotron emission, for example, can be naturally produced by spatially-intermittent magnetic fields \citep[see, e.g.,][]{owen2014, rudnick2022}. Indeed, the filamentary synchrotron morphology in Abell 2256 has, in particular, often been interpreted as evidence for the existence of such magnetic fields \citep{clarke2006, brentjens2008}. In \citetalias{whittingham2024}, we also showed that there was a direct mapping between radio filaments produced in our simulations and the synchrotron-weighted magnetic field strength in projection. However, such structures were not pre-existent in the upstream; they were generated in the downstream through shear and compression. That is to say, in our simulations, magnetic filaments are simply a consequence of including density fluctuations. 

The observation of elongated filaments in face-on relics could help split the degeneracy between `true' one-dimensional filamentary structures and structures seen in projection\footnote{Geometrically, we should expect relics to be produced face-on as well. However, their observation is made difficult by the need to distinguish them from any background radio emission (e.g. radio haloes). Current candidates \citep[see, e.g.,][]{domingue-fernandez2026} are generally at too low resolution to probe for filamentary structures. This situation should improve though with the advent of the SKA telescope.}. However, at present, it is difficult to explain how intermittent magnetic fields can produce periodic morphology; in Sect.~\ref{subsec:shock-corrugation} and Appendix~\ref{appendix:sausage}, we show examples of relics that exhibit `strand' and `knot' features with a remarkably regular spacing of $\sim$500~kpc. There is no known process that can order magnetic filaments independently on this scale.

On a similar theme, it has also been proposed that filamentary forms could reflect variations in the electron acceleration efficiency \citep{wittor2019, rajpurohit2022}. In our own simulations, a varying Mach number at the shock front results in flatter spectra, which substantially changes the resultant intensity\footnote{Stronger shocks also dissipate slightly more thermal energy, which further acts to raise the spectral normalisation. This is, however, a less important effect in our simulations.}. However, as discussed in Sect.~\ref{sec:methodology}, we use a fixed acceleration efficiency of 0.1\%. Although it is generally accepted that stronger shocks are more effective at accelerating electrons, it is still an open question as to how this transition takes place. Several models, though, predict a rise in the acceleration efficiency by $\mathcal{M}\sim5$ to between $\sim$1\% and $\sim$10\% \citep[see, e.g.,][see also Appendix~\ref{appendix:flux-density}]{kang2013, caprioli2014b, ryu2019}. However, this mechanism is still dependent on the Mach number distribution at the shock front, which, once again, we argue is ultimately dependent on density fluctuations. 

Intimately linked to this is the CR electron seed population. It seems clear that re-acceleration is required in order to match the intensity exhibited by observed radio relics \citep[][see also Appendix~\ref{appendix:flux-density}]{pinzke2013, vazza2014, botteon2020}. However, the spatial and energy distribution of the seed population is still unclear. Fossil electrons can be produced in a variety of ways, although the main sources are active galactic nuclei (AGN) jets and through shock acceleration during cosmic structure formation \citepalias[see citations given in Sect.~1 of][]{whittingham2024}. Indeed, these may well work in different scenarios: for example, there is some evidence for radio galaxies actively injecting electrons at close range to radio relics \citep[see, e.g.,][]{johnston2017, botteon2020b}. Overall, however, it is expected that electrons accelerated through structure formation should dominate the energetics\footnote{This is simply because galaxy clusters should dissipate $\sim$$10^{63-64}$ ergs during a merger \citep{markevitch1999}, whilst AGN typically only dissipate $\sim$$10^{61-63}$ ergs over their lifetime \citep{mcnamara2009, vazza2013}. This comparison is of course modified by the volume within which the electrons are injected, hence we discuss both scenarios.}. Whilst there are some statistical predictions for the energy distribution of the fossil population \citep{pinzke2013}, their spatial distribution is currently less well constrained. As a result, it is unclear to what degree these populations will affect relic morphology. Constrained cosmological simulations of the Coma cluster indicate that populations may be patchy \citep{boess2024}. On the other hand, other simulations suggest that sloshing -- the displacement and circulation of gas in response to a large, off-axis gravitational perturbation -- may be highly effective at mixing populations, helping to organise them into arc-like forms \citep{zuhone2021, botteon2024}. Merger- and accretion-driven turbulence is likely similarly effective at dispersing fossil CR electron populations throughout the cluster \citep{perrone2026}. Once again, however, an additional process is required to produce regular structures in relic morphology.

In all cases, simulations will be needed to hone in on how each mechanism affects relic morphology. In future work, we will examine the impact of upstream magnetic filaments and of Mach-number dependent acceleration efficiencies in our shock-tube simulations. We will also use \textsc{Crest} to analyse how seed CR electron populations form in cosmological cluster simulations. Overall, however, we expect such additions to only modify our conclusions, not change them completely.

Indeed, we reiterate that there is now strong evidence for the existence of density fluctuations  in the ICM in both simulations \citep{nagai2011, zhuravleva2013, battaglia2015, angelinelli2021} and observations \citep{simionescu2011, eckhert2015, ghirardini2018}. Moreover, our conclusions in Sect.~\ref{sec:hydro-analysis} are based primarily on well-understood hydrodynamics, whilst those in Sect.~\ref{subsec:radio-emission} are built on standard DSA theory using minimal assumptions. The effects described are therefore a natural consequence of known physics, and must be included in analyses of radio relics even when additional processes are considered. Furthermore, our analyses of mock relic morphology depend only on the statistical properties of the upstream density field; this must not be fine-tuned to produce observed morphological features, in contrast to some of the above mechanisms. Finally, we emphasise the ability of density fluctuations to solve several long-standing puzzles regarding radio relics in a self-consistent manner; it can explain the X-ray~vs.\ radio Mach number discrepancy, the origin of $\upmu$G magnetic field strengths in observations, the spectral index evolution of relics downstream in colour-colour diagrams \citepalias{whittingham2024}, and now the origin of observed morphological features simultaneously.

\section{Conclusions}
\label{sec:conclusions}

Comparatively little is known about gas properties in the outer reaches ($\gtrsim$1~Mpc) of galaxy clusters owing to the low X-ray and SZE surface brightnesses at these distances. Radio relics, however, are bright ($\sim\upmu$Jy~arcsec$^{-2}$ surface brightnesses at 1.4~GHz) and exhibit a wide range of morphological features, which develop as a result of interplay with their environment. In this paper, we show that many of these features are highly sensitive to the power spectra of the upstream density fluctuations. Indeed, we argue that relic morphology can effectively be used to constrain density conditions in the outer ICM of individual clusters.

To show this, we have expanded on the shock-tube simulations first presented in \citetalias{whittingham2024}. These, in turn, were motivated by analysis of the evolution of shocks in cosmological cluster merger simulations. In our expansion, we vary the upstream density field by precisely controlling the parameters of its power spectra. Specifically, we systematically vary the: i) coherence length, ii) coefficient of variation ($\sigma/\mu$; i.e.\ the amplitude), and iii) power-law slope (Figs.~\ref{figure:power-spectra},~\ref{figure:electron-density-pdf} and Table~\ref{tab:simulation_vars}). We post-process each of our simulations with the CR electron spectral solver \textsc{Crest} \citep{winner2019}, modelling advection, DSA with magnetic-obliquity dependence, and cooling via Coulomb, bremsstrahlung, inverse Compton, and synchrotron losses. We generate emission maps from the resultant spectra using the \textsc{Crayon+} code \citep{werhahn2021}. We thus model the emission self-consistently, using the minimum number of tunable parameters.

We find that radio relic morphology is independently sensitive to each of the parameters described above:
\begin{itemize}
    \item \text{Coherence length:} We show that shocks corrugate due to upstream underdensities (Figs.~\ref{figure:density-slices} and~\ref{figure:shock-filaments}). In projection, this leads to `knot' and `double strand' features, as observed in real radio relics. Analysing the Toothbrush and Sausage relics, we find these features have a regular spacing of approximately 500~kpc. We argue that density fluctuations provide the most natural explanation for such spacing. Furthermore, we argue that this spacing is a direct reflection of the coherence length (Fig.~\ref{figure:shock-filaments}, Appendix~\ref{appendix:sausage}), and that this is hence significantly larger than commonly assumed. We additionally demonstrate that relics must have a structural depth of around 1--2$\times$ the coherence length for `double strand' features to be well-formed (Fig.~\ref{figure:shock-filaments}).
\end{itemize}
\begin{itemize}
    \item \text{Amplitude:} We show that a high amplitude of density fluctuations can lead to relic fragmentation (Fig.~\ref{figure:shock-gaps} and Appendix~\ref{appendix:sausage}). This is especially likely for weaker shocks ($\mathcal{M} \lesssim 2$). Additionally, it results in both increased average surface brightness and intensity variations (Figs.~\ref{figure:mach-2-variations} and~\ref{figure:mach-3-variations}, and Appendix~\ref{appendix:flux-density}), caused primarily by a broadened Mach number distribution (Appendix~\ref{appendix:mach-no-dist}). Increasing the amplitude of the fluctuations also produces more extended downstream emission (Figs.~\ref{figure:density-slices}, ~\ref{figure:mach-2-variations}, and~\ref{figure:mach-3-variations}). This provides a natural explanation as to why relics are broader than the idealised DSA expectation. We argue that the varying downstream emission exhibited by the Toothbrush relic is evidence for a varying amplitude of upstream fluctuations.
\end{itemize}
\begin{itemize}    
    \item \text{Power-law slope:} We show that steeper slopes (i.e.\ smoother fluctuations) can result in the formation of large eddies downstream (Fig.~\ref{figure:density-slices}). These, in turn, can produce curved synchrotron filaments (Figs.~\ref{figure:mach-2-variations}, and~\ref{figure:mach-3-variations}). Steepening the power-law slope also results in more synchrotron filaments being formed parallel to the shock front, with larger spacing between them (Figs.~\ref{figure:mach-2-variations}, and~\ref{figure:mach-3-variations}). Finally, it helps to produce more coherent `double strand' features (Figs.~\ref{figure:mock-relic-time-series},~\ref{figure:radio-filaments-vs-frequency}, and Appendix~\ref{appendix:power-law-filaments}). By comparing with observations, and making timescale arguments (Sect.~\ref{subsec:shock-corrugation}), we argue for the existence of steep power-law slopes ($\delta \lesssim -5/3$) in the outskirts of galaxy clusters. 
\end{itemize}

Lastly, in Appendix~\ref{appendix:flux-density} we show which additional factors are required in order for our simulations to match the flux density of observed relics. In particular, we demonstrate the continued need for Fermi I re-acceleration and the likely need for Mach-number dependent acceleration in order to match the brightest relics. 

\begin{acknowledgements}
    The authors thank Kamlesh Rajpurohit for providing data used in Fig.~\ref{figure:shock-filaments} of this paper and Gabriella di Gennaro for providing data used in Fig.~\ref{figure:sausage}. JW acknowledges support by the German Science Foundation (DFG) under grant 444932369. CP and JW acknowledge support by the European Research Council under ERC-AdG grant PICOGAL-101019746. CP and LJ acknowledge support by the DFG Research Unit FOR-5195. PG acknowledges financial support from the European Research Council via the ERC Synergy Grant `ECOGAL' (project ID 855130).
\end{acknowledgements}

\section*{Data Availability}

Movies are available on youtube via URLs indicated in the text.
The data underlying this article will be shared on reasonable request to the corresponding author.

%
   \bibliographystyle{aa} 
   \bibliography{bibliography.bib}

%

\newpage

\begin{appendix}

\section[Appendix A: Impact of the power-law slope on filamentary morphology]{Impact of the power-law slope on filamentary morphology}
\label{appendix:power-law-filaments}

\begin{figure}
    \centering
    \includegraphics[width=\columnwidth]{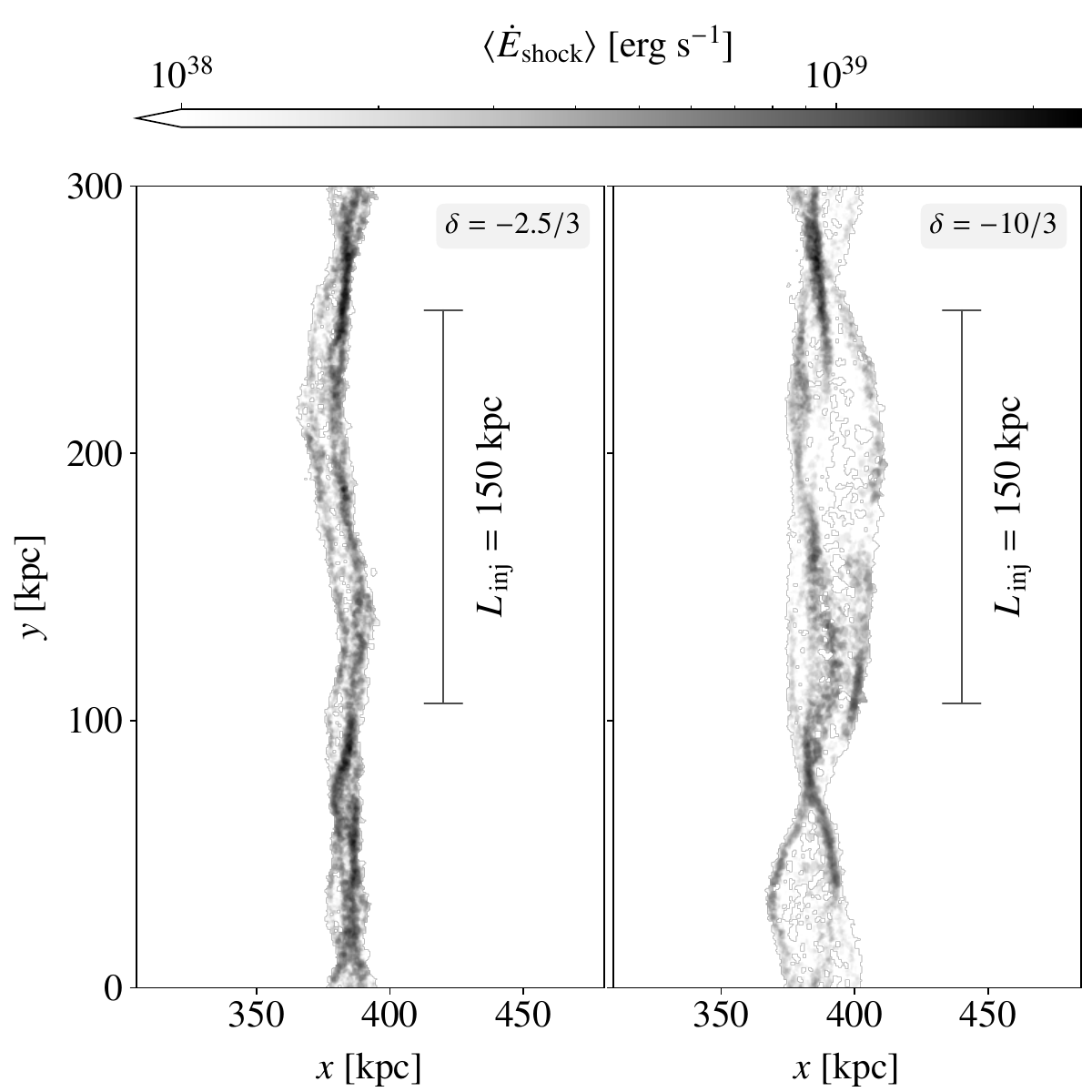}
    \caption{As Fig.~\ref{figure:shock-filaments}, but showing data from our $\delta = - 2.5/3$ and $\delta = - 10/3$ simulations at $t=244$~Myr (left and right panels, respectively). The projection depth is 150~kpc in both cases. Steepening the power-law slope results in smoother shock fronts and better separation between `strand' features.}
    \label{fig:filaments-power-law-slope}
\end{figure}

In Sect.~\ref{subsec:shock-corrugation}, we argued that the `double strand' morphology is influenced by the power-law slope of the upstream density power spectrum. In addition to Sect.~\ref{subsec:radio-filaments}, we support this claim with Fig.~\ref{fig:filaments-power-law-slope}. This figure is analogous to the left-hand side of Fig.~\ref{figure:shock-filaments}, but presents data from our $\delta = - 2.5/3$ and $\delta = - 10/3$ simulations (left and right panels, respectively). This time the `double strand' structure is most clearly developed at $t=244$~Myr. As before, however, the feature is evident also at earlier times. In particular, the $\delta = - 10/3$ simulation shows a persistent corrugation of the shock on the scale of the coherence length over the full runtime.

As previously discussed, steepening the power-law slope results in smoother upstream fluctuations. This, in turn, produces a correspondingly smoother shock front, like the one exhibited by the Toothbrush relic (see Fig.~\ref{figure:shock-filaments}). Additionally, smoother fluctuations allow the shock to propagate farther before encountering a dense clump. This increases the possible separation between individual strands, making it more likely that this feature will be observed (see discussion in Sect.~\ref{subsec:shock-corrugation}).

\section[Appendix B: Applying our analysis to the Sausage relic]{Applying our analysis to the Sausage relic}
\label{appendix:sausage}

\begin{figure*}
    \centering
    \includegraphics[width=2.\columnwidth]{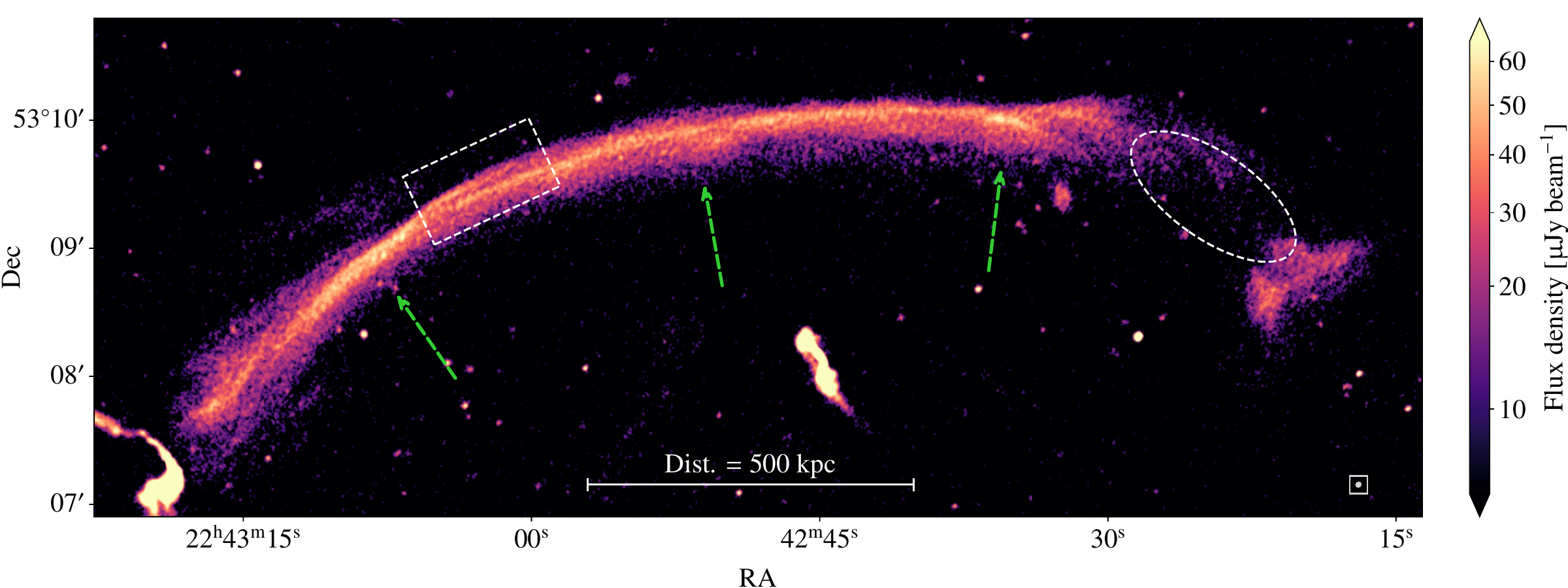}
    \caption{VLA $L$-band 1–2~GHz high-resolution image of the Sausage relic, as first presented in \citet{diGennaro2018}. The beam size is shown in the bottom right corner. Green arrows indicate the location of `knots' along the relic. The dashed white box indicates a radio spur, as evident in Fig.~\ref{figure:mock-relic-time-series}. The dashed white ellipse indicates a potentially missing part of the relic (cf. Fig.~\ref{figure:shock-gaps}). If produced by shock projection, the `knots' along the relic imply a coherence length of $\sim$500~kpc. This is consistent with our analysis of the Toothbrush relic in Fig.~\ref{figure:shock-filaments}, and is difficult to explain through alternative mechanisms.}
    \label{figure:sausage}
\end{figure*} 

In Sect.~\ref{subsec:shock-corrugation}, we showed that density fluctuations lead to shock corrugation and that this, in turn, causes the projected shock surface to appear filamentary. Moreover, we demonstrated that, in projection, this causes the formation of `knots' with spacings that roughly match the coherence length of the upstream density field. As a result, filamentary features in relics may be a direct indication of this scale. Here, we extend our analysis to observations of the Sausage relic  \citep{kocevski2007, vanweeren2010}. Specifically, we use the VLA~$L$-band 1--2~GHz image, as first presented in \citet{diGennaro2018}. We show this image in Fig.~\ref{figure:sausage}. Once again, we have used a power-law scaling of the colour map to better highlight the filamentary features. Unlike in Fig.~\ref{figure:shock-filaments}, however, we have not rotated the image, and hence leave the axes in typical observation units. We have added a representative scale of 500~kpc, assuming the relic is seen edge-on at a redshift of $0.1912$ \citep{vanweeren2010}.

Unlike in the Toothbrush relic, the strands in the Sausage relic are not well-separated. There are several reasons why this may be the case, including projection effects, a more extended downstream, or a less corrugated shock front (see Appendix~\ref{appendix:power-law-filaments}). Nonetheless, some degree of filamentary morphology is still evident, as well as clear `knot' features\footnote{These features should become clearer with the advent of Square Kilometre Array, which will improve resolution by an order of magnitude \citep{braun2019}.}. We indicate the approximate position of these with dashed green arrows. The `knots' tend to be brighter than the surrounding regions. In our picture, this is due to more of the shock front being probed with a single line of sight. Once again, the spacing of the knots is remarkably uniform, measuring approximately $\sim$500~kpc, just as in the case of the Toothbrush relic. This is intriguing as it points towards a process that is able to regulate structures on this scale. Magnetic filaments alone, for example, would be unlikely to self-organise with such regularity (see discussion in Sect.~\ref{sec:discussion}). 

On the right hand side of Fig.~\ref{figure:sausage}, there is a diffuse region of radio emission that is separated from the main relic. This region has similar intensity and spectral index gradation to the rest of the relic \citep[see, in particular, Fig.~3 of][]{vanweeren2010} and so is likely to have a common origin. At the edge of the main relic, there is a region of low-brightness ($\approx15\,\upmu\rmn{Jy}\,\rmn{beam}^{-1}$) emission, which bridges some of the gap. We highlight this with a dashed white ellipse. Moreover, this `bridge' region has a lower average spectral index \citep{vanweeren2010}. This is to be expected in a scenario where this feature forms through fragmentation due to density fluctuations: lower density material results in lower Mach numbers with correspondingly steeper spectral indices and lower intensity emission (see \citetalias{whittingham2024} and Appendix~\ref{appendix:flux-density}). As shown in Sect.~\ref{subsec:shock-fragmenetation}, this can lead to a fragmentation of the shock surface, and hence a lack of emission, even excluding the impact of a critical Mach number. As shown in Fig.~\ref{figure:shock-gaps}, the Mach number typically changes smoothly along the shock surface, but can occasionally jump. This is consistent with the fade-out of emission close to the main relic and the relatively sharp boundary in emission exhibited by the separated region.

Whilst this is a plausible mechanism, we cannot yet rule out alternative mechanisms (see Sect.~\ref{sec:discussion}). In particular, further simulations will be required in order to pin down the impact of an inhomogeneous fossil electron population.

\section[Appendix C: Impact of our models on the Mach number distribution]{Impact of our models on the Mach number distribution}
\label{appendix:mach-no-dist}

\begin{figure}
    \centering
    \includegraphics[width=\columnwidth]{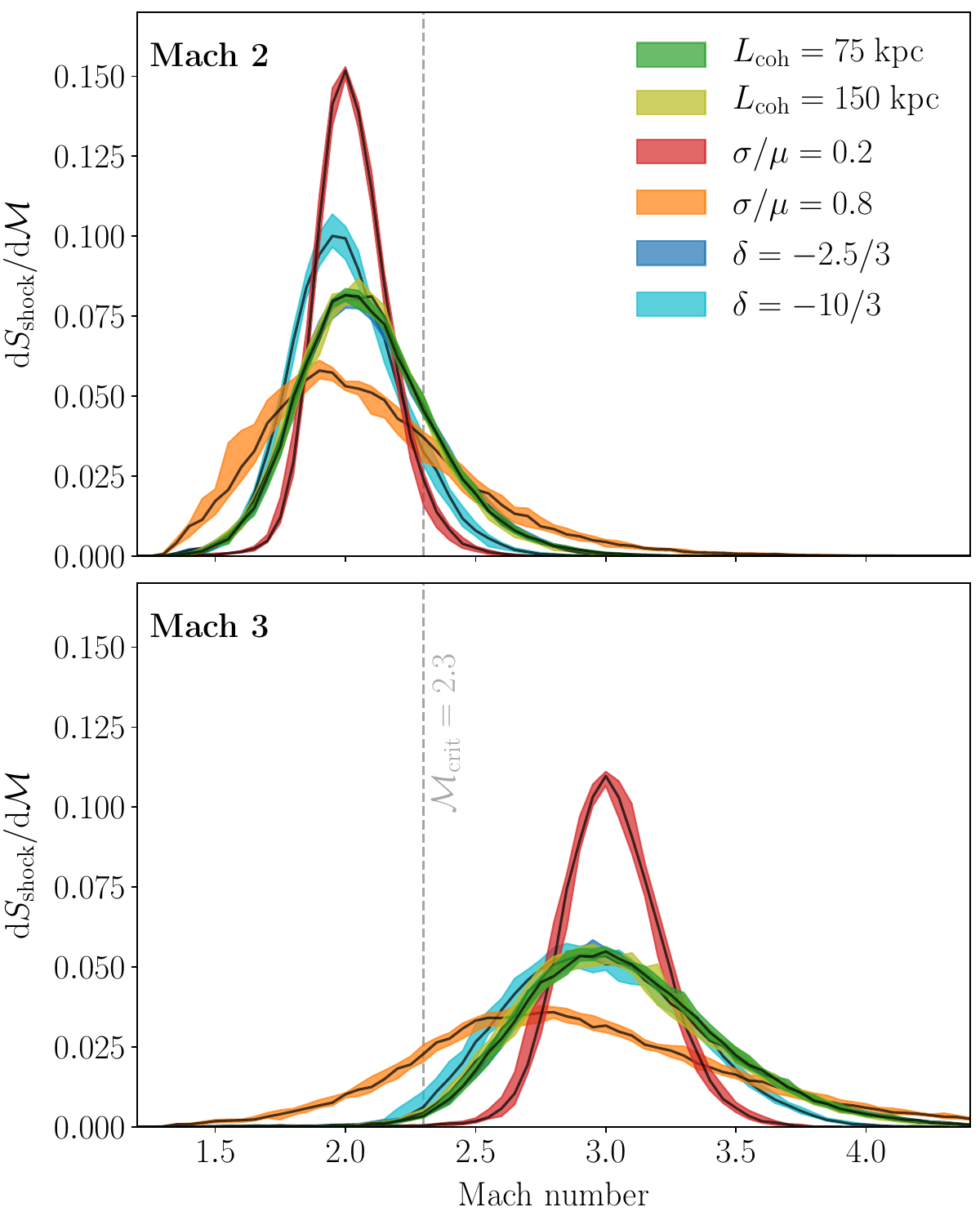}
    \caption{\textit{Top:} Mach number distributions for our $\mathcal{M}=2$ simulations, where cells are weighted by their contribution to the shock surface. Black lines indicate the median taken over all snapshots. Shaded values indicate the interquartile range. The gray, vertical, dashed line marks the critical Mach number above which we inject CR electrons (see Sect.~\ref{sec:methodology}). We additionally allow for a numerical minimum of $\mathcal{M}=1.3$. \textit{Bottom:} As previous, but for our $\mathcal{M}=3$ simulations. The amplitude of the fluctuations has the greatest impact on the distribution, with higher values producing broader distributions.\\
    }
    \label{figure:mach-no-pdf--all-sims}
\end{figure}

In Sect.~\ref{subsec:shock-fragmenetation}, we briefly discussed the formation of a Mach number distribution at the shock front. This phenomena is well-known \citep[see, e.g.,][]{hoeft2011, skillman2013, hong2015, roh2019, wittor2019, dominguez-fernandez2021, wittor2021} and was also investigated in \citetalias{whittingham2024}. Here, we confirm our previous assertion that stronger density fluctuations produce a broader Mach number distribution. In Fig.~\ref{figure:mach-no-pdf--all-sims}, we show the distribution of Mach numbers at the shock front for all simulations. We have weighted each cell by its contribution to the overall shock surface. The black lines represent the medians of the distributions over time, and the shaded regions represent the interquartile ranges. The distributions can each be well-fit by a skew normal function. The vertical, dashed line marks the critical Mach number implemented in our simulations (see Sect.~\ref{sec:methodology}).

It can be seen that by far the largest effect is produced by changing the amplitude of the fluctuations; larger values of $\sigma/\mu$ produce broader distributions, and smaller values make the distribution narrower. This makes sense, as when $\sigma / \mu \rightarrow 0$ we will have no upstream density turbulence and, analytically, a single Mach number as a result. As discussed previously in Sect.~\ref{subsec:shock-fragmenetation}, the upstream gas in our simulations is in pressure equilibrium. Consequently, as the sound speed is $c_\rmn{s}= \sqrt{\gamma_\rmn{a} P/\rho}$, where the adiabatic index of the gas is $\gamma_\rmn{a} \approx 5/3$, density fluctuations also produce sound speed fluctuations. Moreover, the local variations in velocity across the shock front can generally be treated as perturbations as they are small relative to the median shock speed, $\upsilon_\rmn{shock}$. Thus, to first order, the shock strength is simply $\upsilon_\rmn{shock}/c_\rmn{s}(\rho)$. These two facts directly result in the formation of a Mach number distribution, with stronger sound wave fluctuations producing a broader Mach number distribution.

A more subtle effect is produced by steepening the power spectra. For the $\delta=-10/3$ simulations, we observe a slight narrowing of the distribution and a shift to smaller Mach numbers. This effect is more obvious in the Mach~2 run. It is likely that this stems from a biased sampling of overdense regions; steepening the power spectra results in more coherent structures in the density field, meaning that there is less variability in the sound speed. Indeed, in the Mach~3 simulation, the same density field is probed but with a higher average shock velocity. The shock consequently covers more of the upstream initial conditions, thereby better sampling the full density distribution. The result is that the Mach number distribution better matches the other models where $\sigma/\mu = 0.4$ (cf. Table~\ref{tab:simulation_vars}).

\section[Appendix D: Impact of the coherence length on mock emission at 150~MHz]{Impact of the coherence length on mock radio emission at 150~MHz}
\label{appendix:mock-emission-coherence-length}

\begin{figure}  
    \centering
    \small\textbf{Mach~2}\par\medskip
    \includegraphics[width=1.\columnwidth]{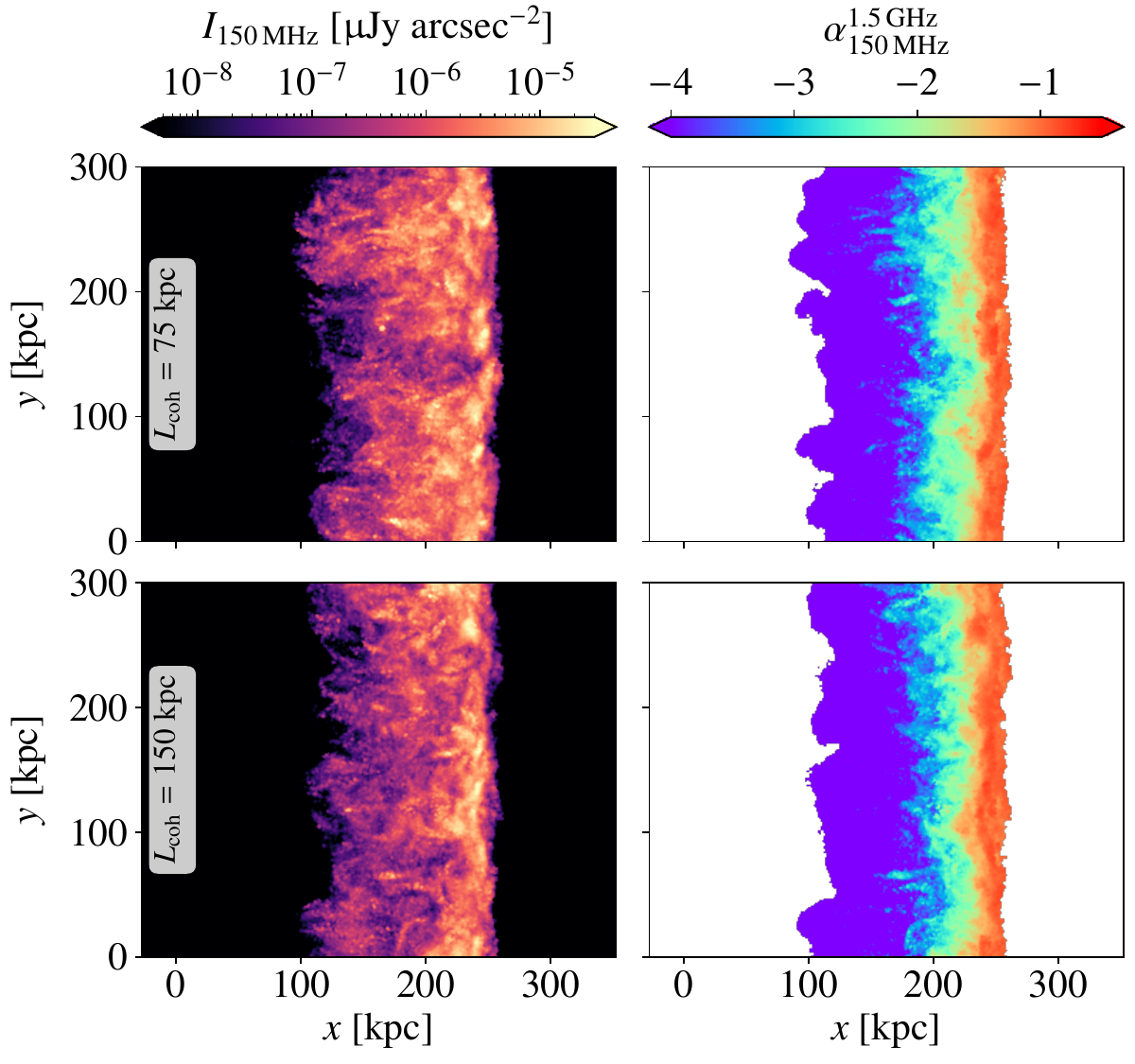}
    \caption{As Fig.~\ref{figure:mach-2-variations}, but showing only the simulations where we varied $L_\rmn{coh}$. Changing the coherence length has only a limited impact at 150~MHz for the sampled parameter space (see Table~\ref{tab:simulation_vars}). An animated version of the figure can be found \href{https://youtu.be/68OLoUN1IZE}{here}.}
    \label{fig:mach-2-l-coh}
\end{figure} 

In Sect.~\ref{subsec:radio-emission}, we analysed the mock radio morphologies produced at 150~MHz by varying the amplitude of the fluctuations and their power-law slope. In Fig.~\ref{fig:mach-2-l-coh}, we vary the coherence length of the fluctuations for the Mach~2 simulations. It can be seen that, out of all the models, changing this parameter has the most limited impact\footnote{This is partially due to the explored parameter space, as discussed at the end of this section.}.
The main effects of increasing the coherence length on these mocks are: i) a slight decrease in average intensity (see spectral flux densities given in Fig.~\ref{figure:flux-density}), and ii) an equally mild decrease in overall downstream extent. We show that these effects are stable over time in the corresponding animated version of the figure \href{https://youtu.be/68OLoUN1IZE}{here}. Both simulations have an essentially identical Mach number distribution at the shock front (see Appendix~\ref{appendix:mach-no-dist}), as this is driven by the amplitude of the fluctuations (i.e.\ $\sigma / \mu$), which are statistically equal. Consequently, almost all variations arise from changes in the Rayleigh-Taylor instability; increasing the coherence length effectively reduces the amplitude of a fluctuation at a fixed wavenumber (see Sect.~\ref{sec:methodology}). This means that the instability is more weekly seeded and takes longer to enter the non-linear regime. As a result, the downstream becomes less extended, and the magnetic field is more weakly amplified, which lowers the intensity of emission.
As we only alter the coherence length by a factor of two, the overall effect is very small.

\begin{figure}   
    \centering
    \small\textbf{Mach~3}\par\medskip
    \includegraphics[width=1.\columnwidth]{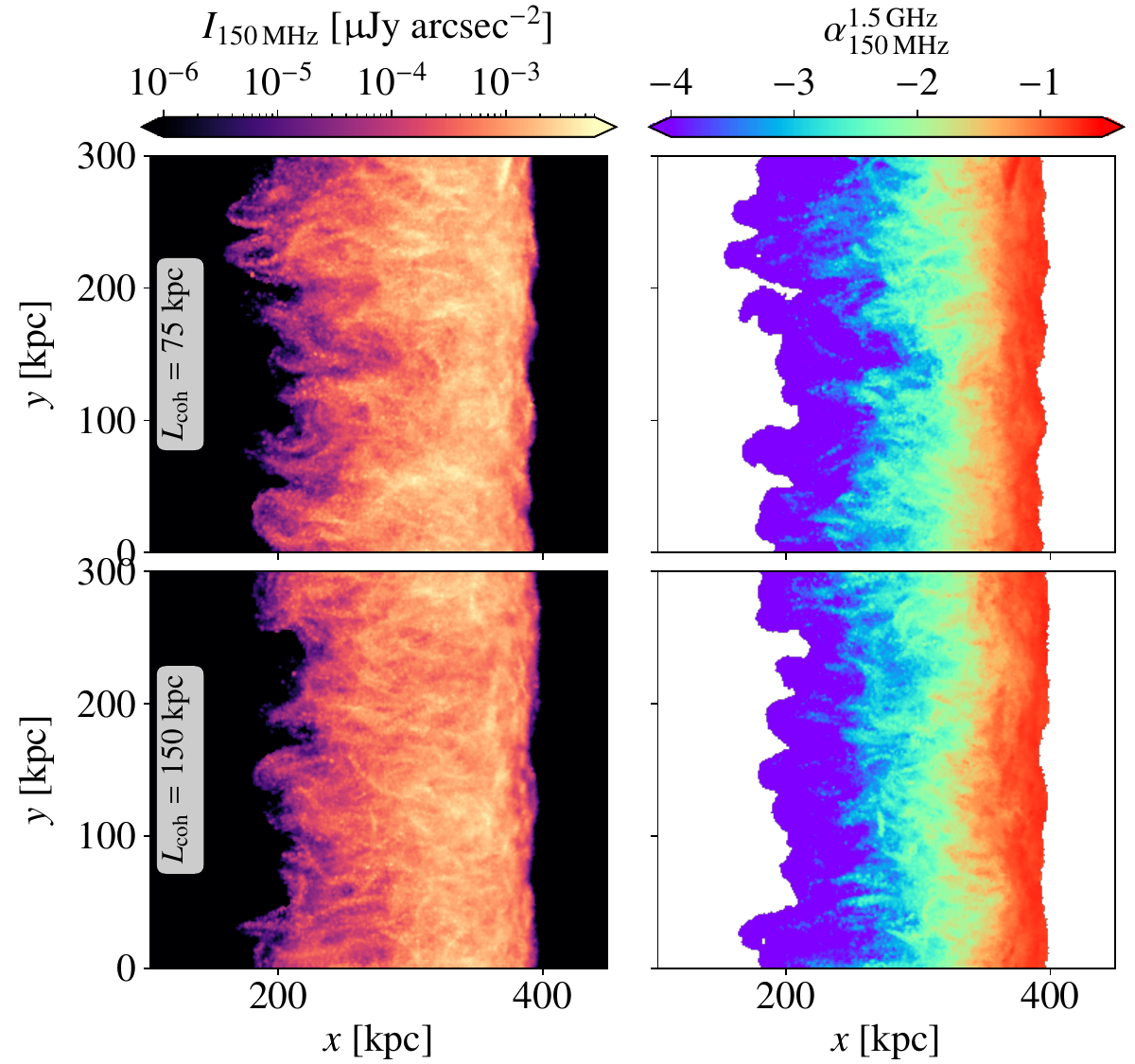}
    \caption{As Fig.~\ref{fig:mach-2-l-coh}, but for the Mach~3 variations. The same conclusion applies. An animated version of the figure can be found \href{https://youtu.be/T036RSr0va8}{here}.} 
    \label{fig:mach-3-l-coh}
\end{figure} 

In Fig~\ref{fig:mach-3-l-coh}, we show the same simulations run with the Mach~3 shocks. The same conclusions continue to hold. As before, the mock relics are brighter than their Mach~2 variants, owing to flatter spectra and more amplified magnetic fields (see discussion in Sect.~\ref{subsec:radio-emission}). The relics are also more extended, owing to the increased downstream advection speeds. As before, this originates from the stronger impact of inertia -- individual clumps being only weakly accelerated by the shock -- and the more developed Rayleigh-Taylor instability, which acts to reinforce the velocity field (see Sect.~\ref{subsec:hydrodynamic-substructure} and \citetalias{whittingham2024}).

In Sect.~\ref{subsec:shock-corrugation}, we argued that the coherence length should set the size of `double strand'-like features in radio morphology. It is thus notable that no such features can be seen in Figs.~\ref{fig:mach-2-l-coh} and~\ref{fig:mach-3-l-coh}. This is primarily a result of the sampled parameter space: for example, our mock intensity maps are shown at low-frequency (150~MHz) and the relics have been developing for the full simulation runtime (244~Myr). Both of these factors result in more extended emission, which is easily able to cover the gap opened up by corrugated shock fronts. In Sect.~\ref{subsec:shock-corrugation}, we also gave evidence that the true coherence length of density fluctuations in the outer ICM is approximately 500~kpc. This is much larger than our fiducial value of 150~kpc, and would additionally help open the gap between thread features (see arguments given in Sect.~\ref{subsec:shock-corrugation}). Finally, and perhaps most importantly, we have also provided evidence in this paper that the spectral slope of the density fluctuations is steep; i.e. $\delta < -5/3$. We show in Sect.~\ref{subsec:radio-filaments}, that threads are evident in our mock radio relics under these conditions, but not when $\delta \geq -5/3$.

\section[Appendix E: Impact of our models on the spectral flux density]{Impact of our models on the spectral flux density}
\label{appendix:flux-density}

\begin{figure}   
    \centering
    \includegraphics[width=1.\columnwidth]{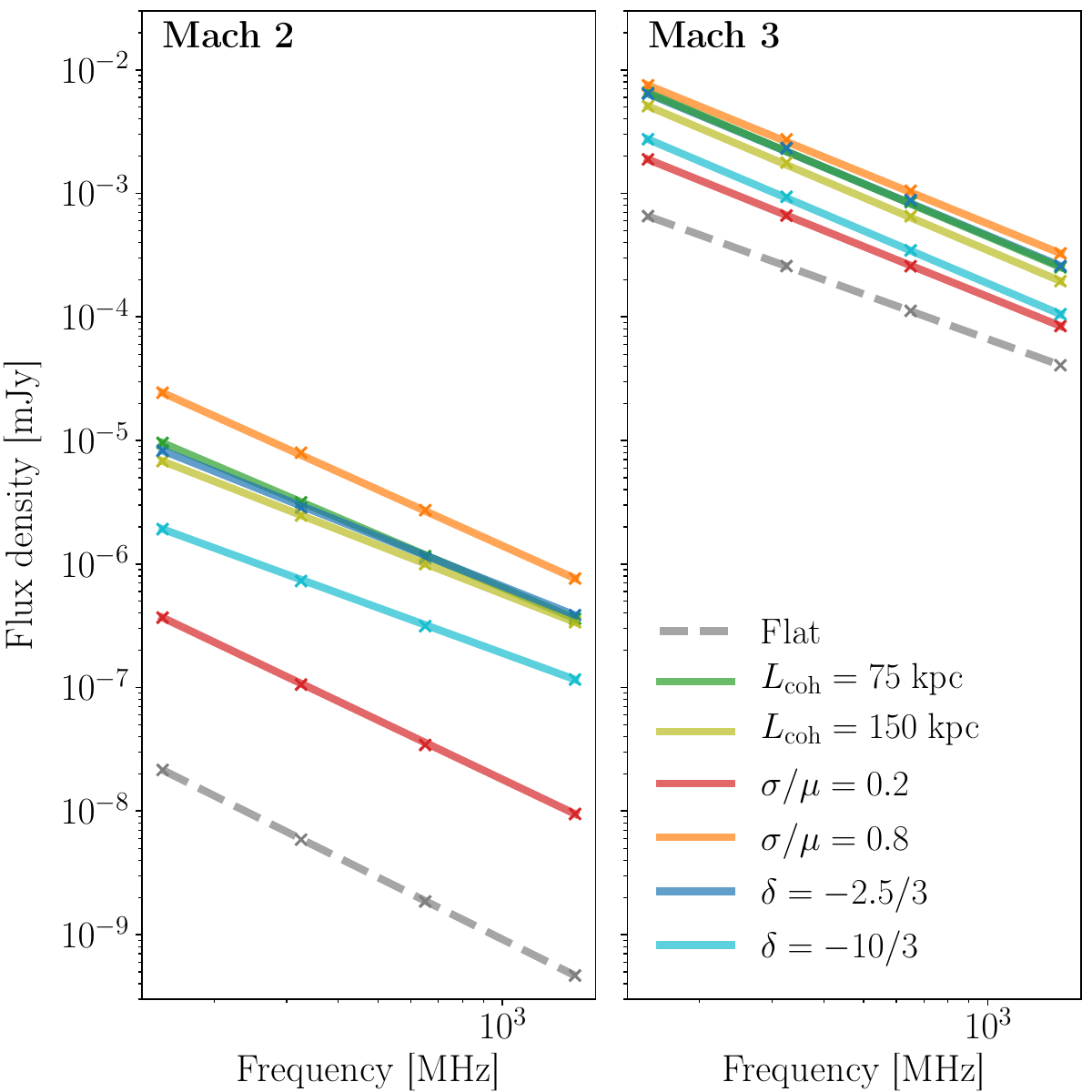}
    \caption{\textit{Left:} Flux density at 150~MHz, 325~MHz, 650~MHz, and 1.5~GHz for our Mach~2 models, measured at $t=244$~Myr. We join the first and last points with a straight line to indicate the lack of deviation from power-law emission. This data can be directly compared with that given in Figs.~\ref{figure:mach-2-variations}, and~\ref{fig:mach-2-l-coh}.
    \textit{Right:} As previous, but for our Mach~3 variations. This data can be directly compared with that given in Figs.~\ref{figure:mach-3-variations}, and~\ref{fig:mach-3-l-coh}. The addition of upstream density fluctuations can substantially increase the spectral flux density. This is especially true for low-Mach number shocks. Nonetheless, Fermi-I re-acceleration of a fossil CR electron population is still required in order to match the observed spectral brightness of real relics (see discussion in Sect.~\ref{subsec:absolute-intensity}).}
    \label{figure:flux-density}
    \vspace{-0.1cm}
\end{figure} 

In Sect.~\ref{subsec:suite-at-150MHz} and Appendix~\ref{appendix:mock-emission-coherence-length}, we showed that the intensity of the downstream emission in our mock relics is highly sensitive to the upstream conditions. Here we quantify that statement by showing the integrated spectral flux density for each model. We calculate this as:
\begin{equation}
    S_\nu = \int_{\Omega_S} I_\nu (\Omega) \mathrm{d}\Omega \equiv \langle I_\nu \rangle_\Omega \,\Omega_S,
    \label{eq:spectral_flux}
\end{equation}
where is $I_\nu$ is the specific intensity at frequency $\nu$ (see Sect.~\ref{subsec:crayon}), and $\Omega_S$ is the solid angle that the relic extends on the sky. On the right-hand side of Eq.~\eqref{eq:spectral_flux}, we have assumed the angle is sufficiently small such that we can use a flat-sky approximation. To aid comparison with observations, we have assumed that our simulated radio relics are at the location of the Toothbrush radio relic, where $\ang{;;1}$ corresponds to a physical scale of 3.64~kpc \citep{rajpurohit2020}. We use emission from the whole box for our calculations. In Fig.~\ref{figure:flux-density}, we show the resultant flux density at 150~MHz, 325~MHz, 650~MHz, and 1.5~GHz for all our models as crosses, with the Mach~2 models shown on the left-hand side and the Mach~3 models shown on the right. We have joined the 150~MHz and 1.5~GHz data points with a straight line to indicate the lack of deviations from the expected power-law emission. It can be seen that all of the models produce integrated emission that is well-described by a single power-law as is the case for real radio relics \citep{rajpurohit2020b, rajpurohit2022}. This result is not trivial, as whilst we inject individual power-law spectra through Eq.~\eqref{eq:spectral-slope}, the integrated spectrum is highly influenced by the fluctuating Mach number distribution and subsequent variable cooling.

Below, we analyse: i) the relative changes in flux density between models; ii) the absolute values reached and the subsequent implication for physics implemented in our simulations; and iii) the realism of our mock spectral slopes.

\subsection{Relative intensity}

It can be seen that in every case, adding density fluctuations produces higher intensity emission than the baseline `Flat' simulation (see Table~\ref{tab:simulation_vars}). The intensity of emission from a given CR electron population is set by the local magnetic field strength and the population's spectral slope (see Eq.~\ref{eq:synchrotron-emissivity}). Shallower spectral slopes are produced by stronger Mach numbers, and hence the normalisations seen in Fig.~\ref{figure:flux-density} are also intrinsically linked to the underlying Mach number distribution, as analysed in Appendix~\ref{appendix:mach-no-dist}. In particular, the flux normalisations generally map to the extent of the high Mach-number tail, as shown in Fig.~\ref{figure:mach-no-pdf--all-sims}.

The boost to the overall flux density is particularly strong in the Mach~2 variations, where the flux density is up to three orders of magnitude higher than the `Flat' run. This is because, at lower Mach numbers, only a small increase in the Mach number is required to produce a significantly higher normalisation at radio-emitting frequencies. In contrast, in the Mach~3 simulations, the difference between the $\sigma/\mu=0.8$ and the `Flat' simulation is little more than a factor of 10. This is because stronger shocks already have relatively flat spectral slope, and the consequent increase in normalisation is limited (see Sect.~4.3 of \citetalias{whittingham2024}). This, in turn, limits the overall ability of density fluctuations to increase relic luminosity up to observed levels.

\subsection{Absolute intensity}
\label{subsec:absolute-intensity}

\renewcommand{\arraystretch}{1.2}
\begin{table}
    \centering
    \caption{Parameters that could help boost the flux density of our simulations to that exhibited by the Toothbrush relic.}
    \begin{tabular}{ c | c }
\makecell{Variable} & \makecell{Flux density boost} \\[2pt] \hline

Shock surface area  
& 10 --- 20 $\,$\tablefootmark{(a, b)} \\[2pt]  

Upstream density 
& 10 $\,$\tablefootmark{(b)} \\[2pt]

Shock velocity   
& 4.25 or 1.25 $\,$\tablefootmark{(b, c)} \\[2pt]

Fermi I re-acceleration  
& $10^3$ --- $10^4$ or 10 $\,$\tablefootmark{(c, d)} \\[2pt]

\makecell{{Mach-number} \\ {dependent acceleration}}  
& 10 --- 100 $\,$\tablefootmark{(e)} \\[2pt]

    \end{tabular}
    \label{tab:flux-density-factors}

    \tablefoot{
    \tablefoottext{a}{Assumes a structural depth between 500~kpc and 1~Mpc.}
    \tablefoottext{b}{See Eq.~\eqref{eq:dissipated-energy-obs}.}
    \tablefoottext{c}{Numbers are relevant for Mach~2 and Mach~3 shocks, respectively.}
    \tablefoottext{d}{See \citet{pinzke2013}}
    \tablefoottext{e}{Based on models tabulated in \citet{boess2023}.}
    In combination, these factors could boost the flux density in our Mach~2 and Mach~3 simulations by $\sim$$10^8$ and $\sim$$10^5$, respectively. This would bring the flux density into alignment with that exhibited by the Toothbrush relic \citep{rajpurohit2020}.
    }

    \vspace{0.2cm}
\end{table}

Observed radio relics reach flux densities between one and a few hundred millijanskys at 1.4~GHz \citep[see, e.g., Table 1 in][]{wittor2021}. For instance, the Toothbrush relic, which is a particularly bright example, has a spectral flux density at 1.4~GHz of $S_\rmn{1.4\,GHz} \approx 330$~mJy. This is substantially higher than any value reached by our models. There are several factors that help explain this discrepancy. We list these in Table~\ref{tab:flux-density-factors}, and discuss them below.

One factor is the difference in dissipated energy rate across the shock surface, as inferred from observations compared to our own simulations. For an order of magnitude estimate, following \citet{botteon2020}, we define this as\footnote{Note, this formula differs from the precise per-cell calculation used in our simulations (see Eq.~\ref{eq:shock-dissipated-energy}).}:
\begin{equation}
\dot{\tilde{E}}_\rmn{diss} = \frac{1}{2} A \rho_1 \upsilon_\mathrm{s}^3,
\label{eq:dissipated-energy-obs}
\end{equation}
where $A$ is the total shock surface, and the other variables keep their previously defined meaning\footnote{We assume a fixed acceleration efficiency here. We consider a variable efficiency later in the section.}. In Fig.~\ref{figure:flux-density} we integrated the mock flux density values over a depth and height of 300~$\times$ 300~kpc (i.e.\ the box dimensions). The Toothbrush relic, in contrast, has a largest linear size of 1.9~Mpc \citep{rajpurohit2020} and, following our analysis in Sect.~\ref{subsec:shock-corrugation}, a likely depth along the line of sight of between 500~kpc and 1~Mpc. Putting aside possible variations in the downstream extent, accounting for this increased size would boost the flux density of our simulated relics by a factor of approximately 10~to~20, assuming a fixed acceleration efficiency.

The initial conditions of our simulations enter in the final two variables of Eq.~\eqref{eq:dissipated-energy-obs}. As discussed in Sect.~\ref{sec:methodology}, the simulations presented in this paper are extensions of those first introduced in \citetalias{whittingham2024}. In turn, the original simulations are based on the case study presented in that paper: specifically, the meeting of a merger shock and an accretion shock in a cosmological cluster merger simulation at a cluster-centric distance of $1.8 \times R_{500} \approx 3$~Mpc. Radio relics have been observed at this distance \citep[see, e.g.,][]{bagchi2011, erler2015}. However, bright relics such as the Toothbrush are typically found significantly further in -- the Toothbrush specifically is located at approximately 1~Mpc from the cluster centre\footnote{The reduced distance does not affect our proposed scenario \citep[see, e.g., study by][]{zhang2020}.} \citep{vanweeren2012}. Here, the gas density is substantially higher. Using X-ray surface brightness profiles extracted across the shock, \citet{botteon2020} calculate the downstream electron number density for the Toothbrush relic to be $n_\rmn{2} = 5.5\times10^{-4}\,\rmn{cm}^{-3}$ and hence, using their reported compression ratio of $C\approx1.37$, the upstream value is $n_\rmn{1}\approx 4\times10^{-4}\,\rmn{cm}^{-3}$. This is a factor of 10 higher than the mean value we use in our simulations (see Fig.~\ref{figure:electron-density-pdf}). Hence, via Eq.~\eqref{eq:dissipated-energy-obs}, this accounts for an additional factor of 10 in flux density.

Through the same method as above, \citet{botteon2020} calculate the downstream temperature to be roughly $9.5 \times 10^7\,\rmn{K}$. Assuming the gas is predominantly hydrodynamic (i.e.\ $\gamma_\mathrm{a, eff} = 5/3$), and using the compression factor as before, this implies an upstream temperature of $7.8 \times 10^7\,\rmn{K}$. For fully-ionised, primordial gas, this produces a sound speed of $1,320\,\rmn{km}\,\rmn{s}^{-1}$. Assuming $\mathcal{M} = 1.23$, as calculated using the Rankine-Hugoniot conditions, the average shock speed is then $\upsilon_\mathrm{s} = 1,620\,\rmn{km}\,\rmn{s}^{-1}$. This increased speed would have a minor effect on our own data, resulting in an increase in the flux density by a factor of $(1,620 / 1,000)^3 \approx 4.25$ and $(1,620 / 1,500)^3 \approx 1.25$ in our Mach~2 and Mach~3 simulations, respectively. Multiplying all these factors together, we find that the shock dissipated energy rate alone accounts for a difference of approximately 100--1000$\times$ in luminosity between our simulated relics and the Toothbrush\footnote{To make these calculations, we have assumed a constant upstream density and a single Mach number. This should be sufficient for order of magnitude estimates but, as shown in this paper and in \citetalias{whittingham2024}, is not accurate beyond this.}. This will naturally be accounted for in future studies when we will attempt to replicate specific examples.

A major additional boost could be realised through Fermi-I re-acceleration. As discussed in Sect.~\ref{sec:methodology}, we have assumed an initially thermal state for our CR electrons. This was chosen for two main reasons: i) the true distribution of non-thermal electrons in clusters is uncertain, and ii) the impact of pre-existing non-thermal electrons should only change the normalisation of the CR electron spectral slope, and not the spectral index\footnote{See caveats discussed in Sect.~\ref{sec:discussion}.} \citep[see, e.g., results and citations in][]{winner2019}, as primarily dealt with in this work. In \citet{pinzke2013}, however, it was shown that Fermi-I re-acceleration can likely boost the spectral normalisation, and hence flux density, by a factor of $10^3$--$10^4$ in Mach~2 shocks and by a factor of 10 in Mach~3 shocks (see their Fig.~9, in particular). These values are also consistent with re-acceleration in so-called multi-shock scenarios, as modelled by \citet{smolinski2023}. If the Toothbrush relic has a mean Mach number of $\mathcal{M} = 1.23$, as quoted previously, it may stand to benefit even more from re-acceleration. Nonetheless, a full numerical simulation will still be required to understand the true impact, especially given the existence of a Mach number distribution at the shock front.

Finally, it is possible that the energy conversion fraction we have implemented is too conservative. As stated in Sect.~\ref{sec:methodology}, we use 0.1\% of the liberated thermal energy. This is common for low-Mach number, quasi-parallel shocks \citep[see, e.g.,][]{gupta2024}. However, it is also well-known that acceleration efficiency increases with Mach number. Indeed, several Mach-number dependent efficiency models now exist (see Sect.~\ref{sec:discussion}). If the higher Mach numbers dominate the radio emission, as is believed to be the case, this could lead to an additional flux density boost of between 10 and 100 \citep[see, e.g., model tabulations in Fig.\ 1 of][]{boess2023}.

Overall, we conclude that, once the above factors are taken into account, the flux density exhibited by the Toothbrush relic can be reached through a DSA scenario. However, based on the above calculations we do not expect to find relics significantly brighter than the Toothbrush.

\subsection{Spectral slope}

The individual spectral slopes exhibited by our mock radio relics vary, with slopes measured in Fig.~\ref{figure:flux-density} ranging from $\alpha^{1.5\,\rmn{GHz}}_{150\,\rmn{MHz}} = -1.6$ to $\alpha^{1.5\,\rmn{GHz}}_{150\,\rmn{MHz}} = -1.2$. This is broadly consistent with the range reported for observed radio relics \citep[see, e.g., Table 1 in][]{wittor2021}. We note, however, that none of our simulated relics have spectral slopes close to -1.1 at the time shown ($t=244$~Myr), whereas such values have been measured for real radio relics \citep[see, e.g.,][]{hoang2018}. Such slopes exist in our simulations only at earlier times.

Moreover, in the Mach~3 simulations it is clear that some of our models exhibit steeper spectral slopes than that of the `Flat' run. This is perhaps unexpected, even given a fluctuating Mach number distribution; the broader Mach number distributions observed in Fig.~\ref{appendix:mach-no-dist} should, naively, always produce flatter slopes. The origin of these unexpectedly steep slopes appears to lie in the interplay between the freshly injected electron population and the already established, cooled component of the integrated spectrum. A detailed investigation of this effect is, however, out of the scope of this paper, and so we leave it to future work.

\end{appendix}

\end{document}